\documentclass[11pt,a4paper,useAMS,usenatbib]{emulateapj}
\usepackage{epsfig}
\usepackage{amsmath}
\usepackage{natbib}
\usepackage{rotating}

\begin{document}

\title{Hot X-ray coronae around massive spiral galaxies: \\  a unique probe of structure formation models}

\author{\'Akos Bogd\'an\altaffilmark{1,4}, William R. Forman\altaffilmark{1}, Mark Vogelsberger\altaffilmark{1}, \\ Herv\'e Bourdin\altaffilmark{2} ,  Debora Sijacki\altaffilmark{1},  Pasquale Mazzotta\altaffilmark{1,2}, Ralph P. Kraft\altaffilmark{1},  \\ Christine Jones\altaffilmark{1}, Marat Gilfanov\altaffilmark{3}, Eugene Churazov\altaffilmark{3}, and Laurence P. David\altaffilmark{1}}
\affil{\altaffilmark{1}Harvard Smithsonian Center for Astrophysics, 60 Garden Street, Cambridge, MA 02138, USA; abogdan@cfa.harvard.edu}
\affil{\altaffilmark{2}Dipartimento di Fisica, Universit\`a degli Studi di Roma ``Tor Vergata'', via della Ricerca Scientifica 1, 00133 Roma, Italy}
\affil{\altaffilmark{3}Max-Planck-Institut f\"ur Astrophysik, Karl-Schwarzschild-str. 1, 85748 Garching, Germany}

\email{$^4$Einstein Fellow}

\shorttitle{HOT X-RAY CORONAE AROUND MASSIVE SPIRALS}
\shortauthors{BOGD\'AN ET AL.}

\begin{abstract}
Luminous X-ray gas coronae in the dark matter halos of massive spiral galaxies are a fundamental prediction of structure formation models, yet only a few such coronae have been detected so far. In this paper, we study the hot X-ray coronae beyond the optical disks of two ``normal'' massive spirals, NGC1961 and NGC6753. Based on \textit{XMM-Newton} X-ray observations,  hot gaseous emission is detected to $\sim$$60$ kpc -- well beyond their optical radii. The hot gas has a best-fit temperature of $kT\sim0.6$ keV and an abundance of $\sim$$0.1$ Solar, and exhibits a fairly uniform distribution, suggesting that the quasi-static gas resides in hydrostatic equilibrium in the potential well of the galaxies. The bolometric luminosity of the gas in the $(0.05-0.15)r_{200}$ region ($r_{200}$ is the virial radius) is $\sim$$6\times10^{40}\ \rm{erg\ s^{-1}}$ for both galaxies. The baryon mass fractions of NGC1961 and NGC6753 are $f_{\rm{b,NGC1961}}\sim0.11$ and $f_{\rm{b,NGC6753}}\sim0.09$, which values fall short of the cosmic baryon fraction. The hot coronae around NGC1961 and NGC6753 offer an excellent basis to probe structure formation simulations. To this end, the observations are confronted with the moving mesh code \textsc{arepo} and the smoothed particle hydrodynamics code \textsc{gadget}. Although neither model gives a perfect description, the observed luminosities, gas masses, and abundances favor the \textsc{arepo} code. Moreover, the shape and the normalization of the observed density profiles are better reproduced by \textsc{arepo} within $\sim$$0.5r_{200}$. However, neither model incorporates efficient feedback from supermassive black holes or supernovae, which could alter the simulated properties of the X-ray coronae. With the further advance of numerical models, the present observations will be essential in constraining the feedback effects in structure formation simulations.
\end{abstract}


\section{Introduction}
\label{sec:introduction}
The presence of hot gaseous coronae in dark matter halos of massive galaxies is a fundamental prediction of structure formation models \citep{white78}. This gas collapse scenario in a cold dark matter dominated Universe was applied by \citet{white91},  who predicted that dark matter halos -- surrounding all types of present epoch massive galaxies -- should have an associated X-ray luminous corona. 

X-ray luminous, hot gaseous coronae are nearly ubiquitous for massive early-type galaxies. First discovered with the \textit{Einstein Observatory} \citep{forman85}, these coronae have been extensively studied \citep[e.g.][]{sarazin01,jones02,mathews03,bogdan11b}. Early-type galaxies exhibit at least three X-ray emitting components. (1) A well-studied population of low-mass X-ray binaries (LMXBs), which add a notable contribution to the total X-ray emission of galaxies
\citep{gilfanov04,zhang12}; (2) emission from active binaries (ABs) and cataclysmic variables (CVs), whose total emission is proportional to the stellar mass of the galaxy \citep[e.g.][]{revnivtsev08,bogdan11b}, and (3) thermal, sub-keV gas, produced from the galaxy’s own constituent stars or from infall of group/cluster gas in central bright cluster galaxies. For massive early-type galaxies, it is difficult to detect the more extended atmospheres of hot gas, since they either have bright coronae from stellar mass loss and/or lie in larger dark matter halos –- groups and clusters –- that have their own gaseous atmospheres. Thus, elliptical galaxies are not suitable for probing the luminous X-ray gas predicted around individual galaxies.

\begin{table*}
\caption{The properties of the sample galaxies.}
\begin{minipage}{18.5cm}
\renewcommand{\arraystretch}{1.3}
\centering
\begin{tabular}{c c c c c c c c c c c}
\hline 
Name &  Distance & $1\arcmin$ scale  &  $N_{H}$ & $L_{K} $    &  $M_{\star}/L_{K} $            &  $M_{\star}$ & SFR & Morph.    & $r_{200}$ & $ T_{\mathrm{obs}} $             \\ 
   &   (Mpc)    & (kpc) & (cm$^{-2}$) &   ($\rm{L_{K,\odot}}$) & ($\rm{M_{\odot}/L_{K,\odot}}$) & ($\rm{M_{\odot}}$) &($\rm{M_{\odot} \ yr^{-1}} $) &  type       & (kpc) & (ks)                   \\ 
     &   (1)    &            (2)           &   (3)      &     (4)     &      (5)                 &   (6) &  (7)                  &   (8)     & (9)    &  (10)           \\
\hline 
NGC1961  & $ 55.8  $ & 16.23 & $  8.4 \times 10^{20}$ & $ 5.4 \times 10^{11}$ & 0.78 & $ 4.2 \times 10^{11}$  &  15.5 & SAB(rs)c & 470 & $ 73.7 $ \\
NGC6753  &  $ 43.6 $ &  12.67 & $  5.5 \times 10^{20}$ & $  3.9 \times 10^{11}$& 0.81 &  $ 3.2 \times 10^{11}$ &  11.8 & (R)SA(r)b & 440 &$ 73.9 $  \\
\hline \\
\end{tabular} 
\end{minipage}
\textit{Note.} Columns are as follows. (1) Distance taken from NED (http://nedwww.ipac.caltech.edu/). (2) $1\arcmin$ scale at the applied distance. (3) Galactic absorption \citep{dickey90}. (4) Total K-band luminosity. (5) K-band mass-to-light ratios computed from \citet{bell03} using the $B-V$ color indices of galaxies \citep{devaucouleurs91}. (6) Total stellar mass based on the K-band luminosity and the K-band mass-to-light ratios. (7) Star formation rate computed from the \textit{IRAS} $60 \ \rm{\mu m}$ and $100 \ \rm{\mu m}$ flux, details are given in Section \ref{sec:fir}. (8) Morphological type, taken from NED. (9) Virial radius of the galaxies, estimated from the maximum rotation velocity. (10) Total \textit{XMM-Newton} exposure time. \\

\label{tab:list1}
\end{table*}

However, late-type, spiral galaxies, also are predicted to have luminous X-ray coronae and can test galaxy formation scenarios \citep{white91}. A major advantage of spirals is their location: whereas massive ellipticals lie either in the center of galaxy groups and clusters and/or in rich environments, it is possible to find relatively isolated and quiescent massive spiral galaxies.  Starburst galaxies or galaxies undergoing mergers are not suitable to characterize the extraplanar X-ray emission.  Thanks to UV, far-infrared, and H$\alpha$ imaging, large star-formation rates, hence starburst galaxies, can be identified, and merging galaxies can be recognized based on their optical appearance. Thus, isolated massive spiral galaxies with moderately low star formation rates and relatively undisturbed morphology are the ideal targets for exploring the extraplanar X-ray coronae, thereby testing structure formation models.  

The observed X-ray emission from spiral galaxies consists of multiple components: (1) In addition to LMXBs \citep{gilfanov04}, bright high-mass X-ray binaries (HMXBs), associated with star-formation, are located in the star-forming regions \citep{grimm03}; (2) Also an important contributor to the X-ray flux is the collective emission of faint compact X-ray sources, which contribute to the unresolved, diffuse component. The population of ABs and CVs follow the stellar light distribution \citep{revnivtsev08}, while the emission from young stars and young stellar objects is proportional with the star-formation rates of the host galaxy \citep{bogdan11b}. (3) Additionally, spiral galaxies host at least moderate amounts of ionized gas with sub-keV temperatures \citep{bogdan11b}. These components, enumerated above, provide the bulk of the observed emission within the optical extent of spiral galaxies and are very difficult to model accurately as an integrated component to detect the faint outer corona projected onto this complex region. Therefore, to probe the structure formation models, the hot X-ray coronae must be explored \textit{beyond the optical extent} of spiral galaxies.

The primary goals of the present paper are twofold. First, we aim  to detect and characterize the hot X-ray coronae beyond the optical disks of two massive spiral galaxies, NGC1961 and NGC6753, based on moderately deep \textit{XMM-Newton} X-ray observations. Second,  we aim to confront the observed physical properties of the X-ray coronae with those predicted by modern structure formation models.

\begin{figure*}[t]
  \begin{center}
    \leavevmode
      \epsfxsize=8.7cm\epsfbox{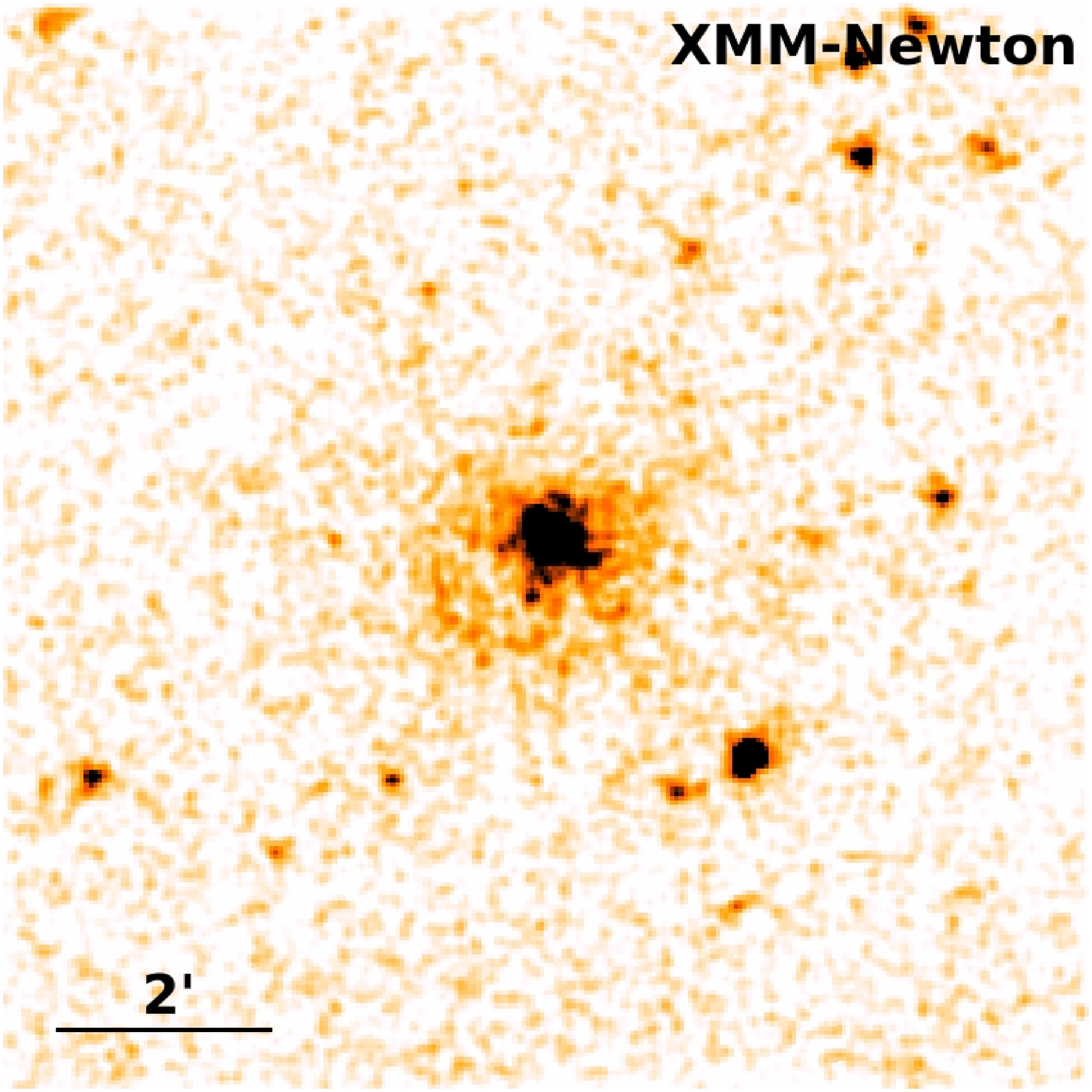}
\hspace{0.2cm} 
      \epsfxsize=8.7cm\epsfbox{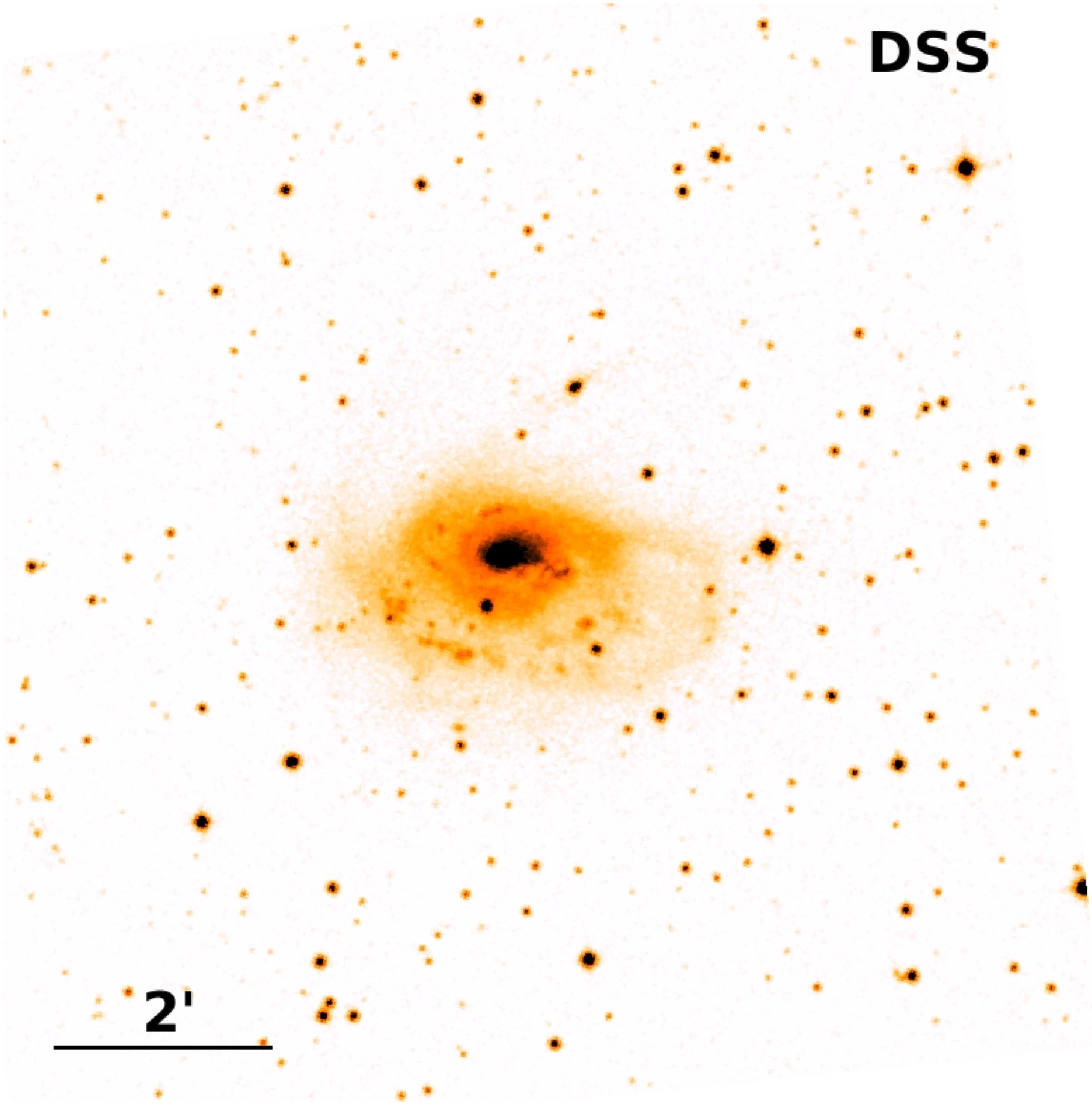}
      \caption{\textit{Left:} $0.3-2$ keV band raw \textit{XMM-Newton} image of a $10\arcmin \times 10\arcmin$ region ($162\times162$ kpc) around NGC1961. The net exposure time of the galaxy after flare filtering is $31.0$, $34.1$, and $21.3$ ks for  MOS1, MOS2, and PN, respectively. The bright X-ray sources beyond the optical radius are cosmic X-ray background sources. \textit{Right:}  DSS R-band image of the same region. The somewhat disturbed morphology may be the result of a recent minor merger \citep{combes09}.}
\vspace{0.5cm}
     \label{fig:ngc1961}
  \end{center}
\end{figure*}

\begin{figure*}
  \begin{center}
    \leavevmode
      \epsfxsize=8.7cm\epsfbox{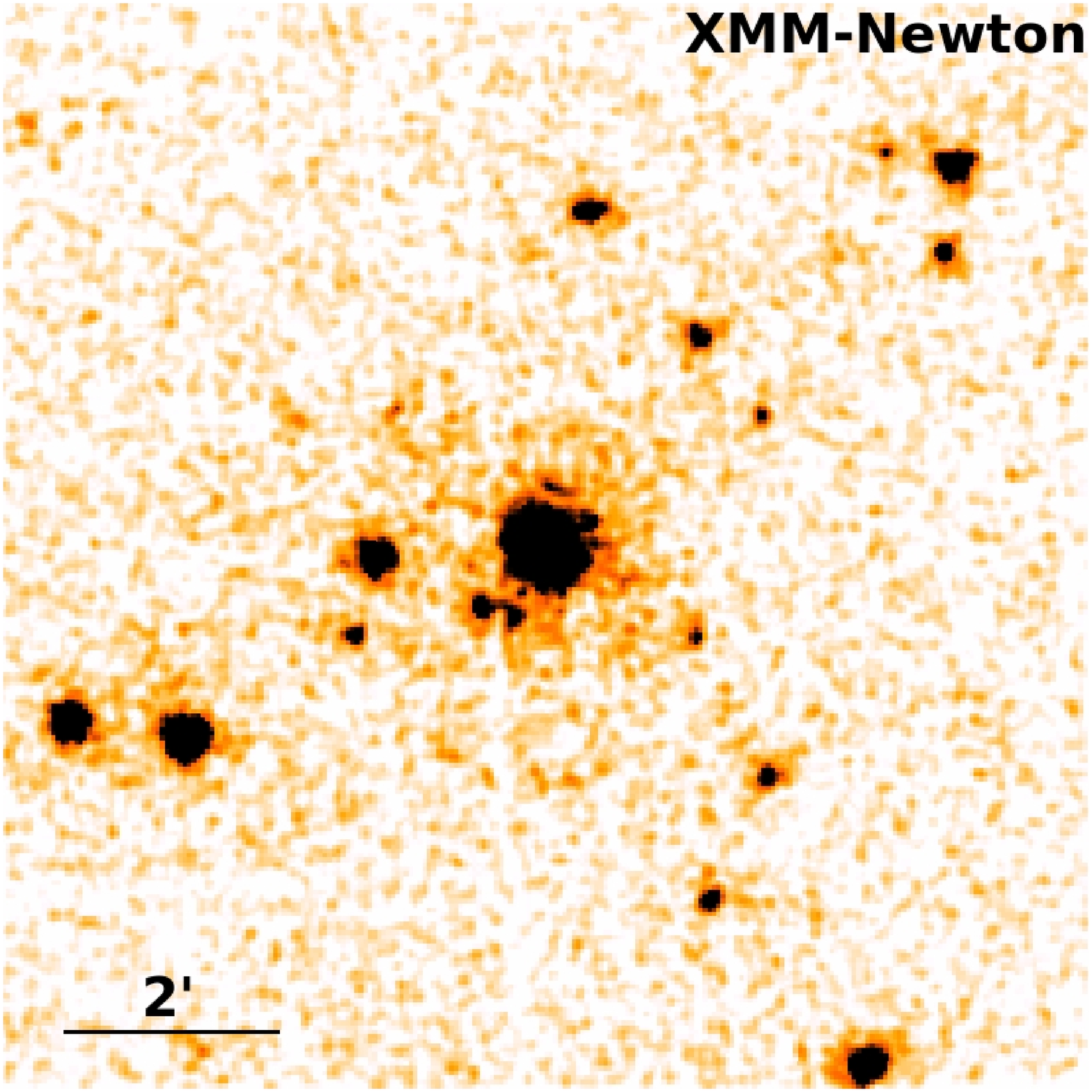}
\hspace{0.2cm} 
      \epsfxsize=8.7cm\epsfbox{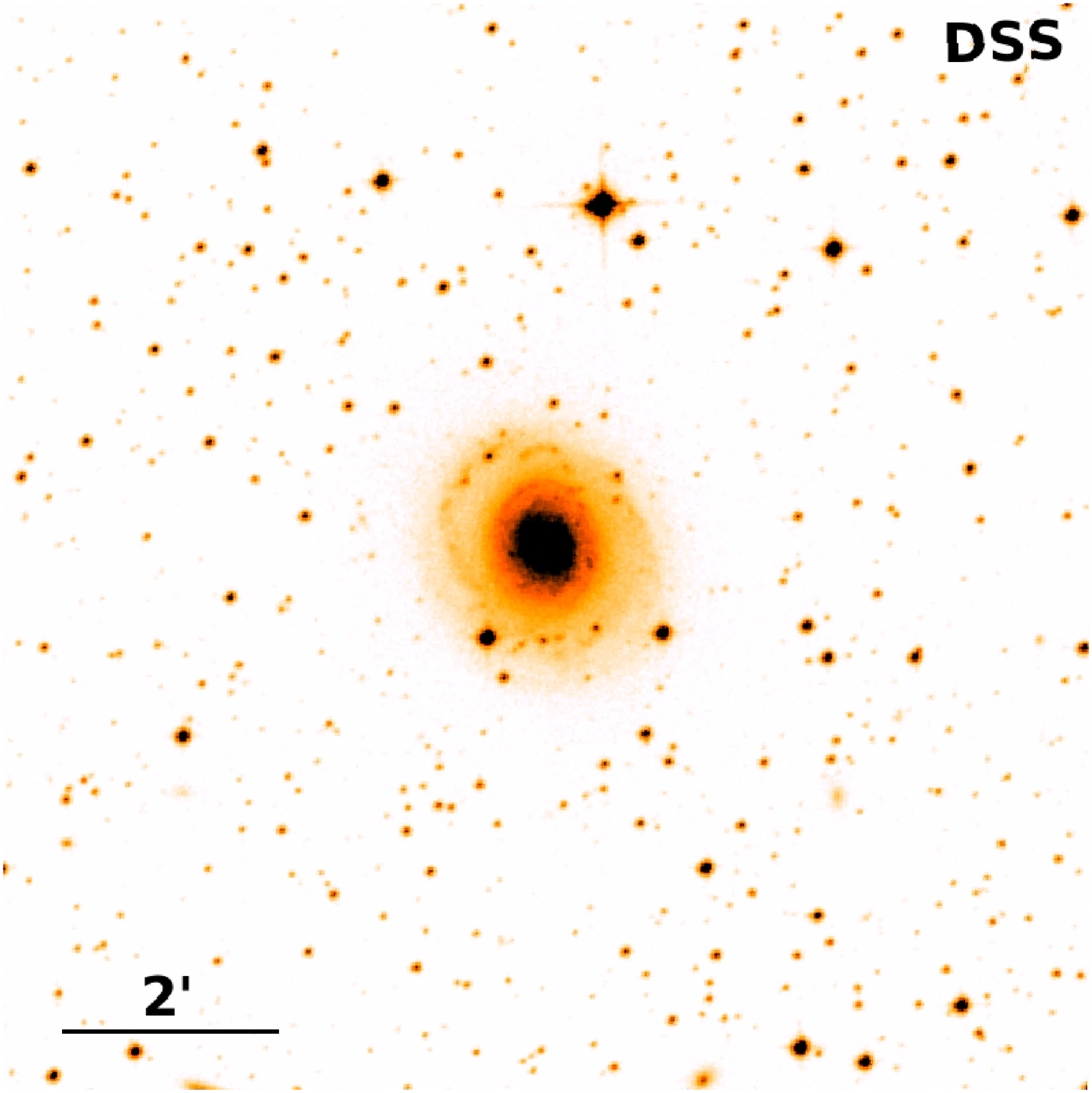}
      \caption{\textit{Left:} $0.3-2$ keV band raw \textit{XMM-Newton} image of a $10\arcmin \times 10\arcmin$ region ($127\times127$ kpc) around NGC6753. The net exposure time of the galaxy after flare filtering is $58.7$, $59.2$, and $27.9$ ks for  MOS1, MOS2, and PN, respectively. The bright X-ray sources beyond the optical radius are cosmic X-ray background sources. \textit{Right:}  DSS R-band image of the same region. The symmetrical morphology of the galaxy indicates its undisturbed nature.}
\vspace{0.5cm}
     \label{fig:ngc6753}
  \end{center}
\end{figure*}

The first analytic analysis of galaxy formation in the cold dark matter cosmogony was presented by \citet{white91}, who suggested that X-ray coronae around massive galaxies should be ubiquitous. For example, their calculations argued that a halo, characterized by $300 \ \rm{km \ s^{-1}}$ circular velocity, would have an X-ray luminosity of $3\times10^{42} \ \rm{erg \ s^{-1}}$ and a gas temperature of about $0.3$ keV. However, the luminosities predicted by \citet{white91} are too high, because they neglected gas ejection and assumed that the hot gas follows the dark matter distribution. The hydrodynamical simulations of \citet{toft02} predict X-ray coronae  about two orders of magnitude fainter than the original \citet{white91} predictions.  The lower predicted luminosities are due to the absence of efficient feedback, which resulted in very massive stellar components, consequently much less massive and less luminous X-ray coronae.  Most recently, \citet{crain10} studied the properties of rotationally supported systems with an extensive simulation of disk-like galaxies. This simulation,  carried out in a large cosmological volume, incorporates efficient  feedback from supernovae that prevents the conversion of halo gas into stars. The supernova feedback also alters the hot gas entropy profile by ejecting a fraction of the baryons to larger radii, and by preferentially ejecting the lowest entropy gas to suppress the central density profile.

The paper is structured as follows. In Section 2 we describe the previously available sample of spiral galaxies and introduce the presently studied NGC1961 and NGC6753. In Section 3 we describe the reduction of the X-ray and infrared data. The results, namely images, surface brightness profiles, and spectra, are presented in Section 4. The observed characteristics of the X-ray coronae are confronted with structure formation simulations in Section 5. We discuss our results in Section 6. We summarize in Section 7.

\section{Previous observations and \\ the present sample of spiral galaxies}
\subsection{Previous observations of spiral galaxies}
Prior to the \textit{XMM-Newton} observations of NGC1961 and NGC6753, the data available for studying the extended X-ray coronae of spiral galaxies was very scarce for the following three reasons. First, most studied spiral galaxies have stellar masses of $M_{\star} \lesssim 3\times 10^{11} \ \rm{M_{\odot}}$, implying rather faint X-ray coronae beyond their optical extent (Section \ref{sec:luminosity}).  Therefore, the predicted faint X-ray coronae could not be detected: only upper limits are obtained, which means that the gaseous emission cannot be characterized. Examples for such spiral galaxies are those studied by \citet{benson00} based on \textit{ROSAT} PSPC data, or the galaxies investigated by \citet{rasmussen09} based on \textit{Chandra} observations. Second, most  galaxies with detected X-ray coronae are \textit{not normal} spiral galaxies, but they are either starburst/post starburst systems, active galactic nuclei (AGN), or spirals whose cold interstellar media have been heated and sometimes stripped by the interaction with the intracluster gas. Therefore, the starburst galaxies, including the famous M82 and NGC253, studied by \citet{strickland04a}, or the galaxies located in rich cluster environment in the \citet{sun07} sample cannot be used to probe structure formation models. Third, the extra-planar X-ray emission detected in a number of normal, edge-on spiral galaxies is not extended to large radii but lies close to the disk \citep[e.g.][]{wang05}. The  extraplanar X-ray emission in these spirals is not likely that arising from an extended corona of infalling gas, but it is most likely a supernova (SN) driven outflow, as discussed by \citet{david06} for low luminosity early-type galaxies and as observed in the bulge of M31 \citep{li07,bogdan08}. Indeed, detailed studies of the extraplanar emission demonstrate that it is confined to within $10$ kpc around NGC5775 \citep{li08} and $23$ kpc around  M104 \citep{li11}, and can be interpreted as gas driven up from the disk.

The first luminous X-ray coronae around massive spiral galaxies have only recently been detected. Most recently, \citet{bogdan13} studied the massive spiral galaxy NGC266 based on a pointed \textit{ROSAT} observation, and detected a luminous X-ray corona extending to at least 70 kpc. Adopting realistic gas temperatures and abundances, \citet{bogdan13} estimated that in the $(0.05-0.15)r_{200}$ region (where $r_{200}$ is the virial radius) the bolometric X-ray luminosity of the hot gas is $(4.3\pm0.8)\times10^{40} \ \rm{erg \ s^{-1}}$. Based on a $130$ ks \textit{Chandra} observation of NGC1961, \citet{anderson11} detected a luminous X-ray corona with the hot gas extending to about $40$ kpc. They measured the spectrum of the integrated emission within 40 kpc to be $kT=0.60^{+0.10}_{-0.09}$ keV temperature  consistent with the expectations for a hot corona. However, the relatively few detected counts did not allow the measurement of the elemental abundances. Using \textit{XMM-Newton} observations,  \citet{dai11} reported the presence of a hot gaseous corona out to $80$ kpc from the center of UGC12591, a rapidly rotating S0/a galaxy. The spectral properties of the gas were measured in the central $\sim24$ kpc region of the galaxy, resulting in a best-fit temperature of $kT=0.60\pm0.03$ keV.

\begin{table*}
\caption{The list of analyzed \textit{XMM-Newton} observations.}
\begin{minipage}{18cm}
\renewcommand{\arraystretch}{1.3}
\centering
\begin{tabular}{c c c c c c c }
\hline 
Galaxy & \textit{XMM-Newton} & Centre  &  MOS1 clean  & MOS2 clean  & PN clean & Observation  \\ 
& Obs ID & coordinates & exposure time (ks) &  exposure time (ks) & exposure time (ks) & Date \\
\hline
NGC1961 & 0673170101 & 05h42m04.60s +69d22'42.0'' & 16.0 & 19.1 & 12.7 & 2011 Aug 31 \\
NGC1961 & 0673170301 & 05h42m04.60s +69d22'42.0'' & 15.0 & 15.0 & 8.6 & 2011 Sep 14 \\
NGC6753 & 0673170201 & 19h11m23.60s $-$57d02'58.0'' & 58.7 & 59.2 & 27.9 & 2012 Apr 21 \\

\hline \\
\end{tabular} 
\end{minipage}

\label{tab:list2}
\end{table*}

\subsection{Spiral galaxies analyzed in this paper}
\label{sec:galaxies}
To detect and characterize hot X-ray coronae with present-day X-ray telescopes, the targeted galaxies must fulfill the following three  criteria. First, the selected spiral galaxies must be massive. More massive galaxies are predicted to have more luminous X-ray coronae, thereby facilitating detections with high signal-to-noise ratios. Second, relatively relaxed galaxies are required in a ``clean'' environment, that is they should not be starburst systems, should not be undergoing an interaction or merger, should not host a significant AGN, and should not be located in a rich galaxy cluster. Finally, the radial range of  $25-100$ kpc should fit in the \textit{XMM-Newton} field-of-view (FOV). The galaxies, NGC1961 and NGC6753 fulfill all these criteria. They are fairly massive ($M_{\star}> 3\times 10^{11} \ \rm{M_{\odot}}$), are quiescent, and are located in sufficiently distant poor galaxy groups \citep{fouque92}. Thus, NGC1961 and NGC6753 are excellent targets to test galaxy formation scenarios. Their major physical properties are listed in Table \ref{tab:list1}. 

To estimate the virial mass ($M_{200}$) of the sample galaxies, we use the baryonic Tully-Fisher relation for the cold dark matter cosmogony, $M_{200} \propto V_{\rm{max}}^{3.23}$, where $V_{\rm{max}}$ is the maximum rotational velocity of the galaxies. The $V_{\rm{max}}$ values are taken from the HyperLeda\footnote{http://leda.univ-lyon1.fr} catalog \citep{paturel03}, which gives $V_{\rm{max}}=447.9\pm14.6 \ \rm{km \ s^{-1}}$ for NGC1961 and $V_{\rm{max}}=395.0\pm21.1  \ \rm{km \ s^{-1}}$ for NGC6753. Using results of numerical simulations \citep[][Vogelsberger et al., in prep.]{navarro97}, we estimate virial masses of $\sim$$1.2\times10^{13} \ \rm{M_{\odot}}$ and $\sim$$1.0\times10^{13} \ \rm{M_{\odot}}$ for NGC1961 and NGC6753, respectively. From the virial masses we compute the virial radii ($r_{200}$) of the galaxies using:
\begin{eqnarray}
M_{200} = 200 \rho_{\rm{crit}} \frac{4 \pi}{3} r_{200}^{3} \ ,
\end{eqnarray}
where $\rho_{\rm{crit}}$ is the critical density of the Universe. We thus estimate that the virial radii of NGC1961 and NGC6753 are $r_{200}\sim470$ kpc and $r_{200}\sim440$ kpc, respectively. 

We emphasize that the applied method serves as a crude estimate of  $M_{200}$ (hence $r_{200}$). Precisely determining these quantitites is a rather convoluted task, which is beyond the scope of this paper. We note that the main application of $M_{200}$ is to determine $r_{200}$. However, as is clear from equation (1), $r_{200}$ is relatively weakly dependent on the virial mass, hence it is much less affected by the uncertainties associated with the determination of $M_{200}$. Thus, the conclusions of this paper remain completely valid, even if the ``real''  virial mass/radius is somewhat different than the quoted values.

\begin{figure*}[t]
  \begin{center}
    \leavevmode
      \epsfxsize=5in\epsfbox{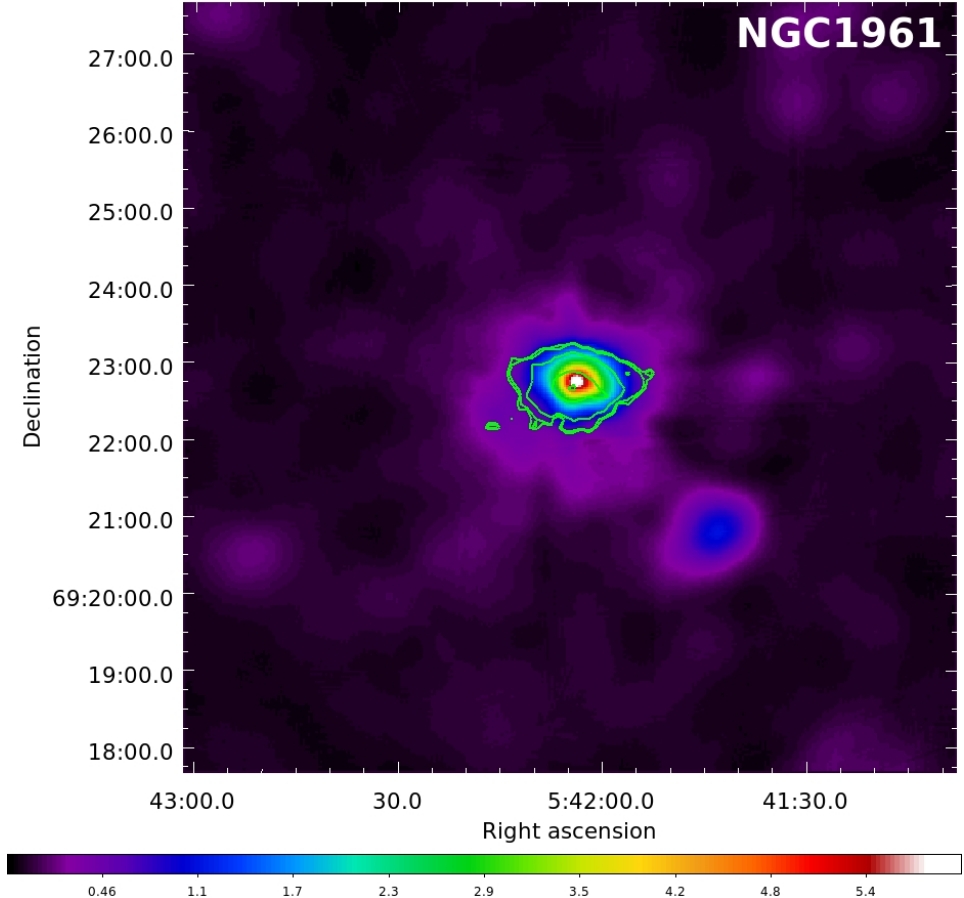}
      \caption{Denoised surface brightness image of NGC1961 based on the combined \textit{XMM-Newton} EPIC datasets. The side length of the image is $10\arcmin$, which corresponds to $162.3$ kpc at the distance of the galaxy. The image reveals large-scale extended emission surrounding the galaxy originating from the hot X-ray gas. Note that point sources are not excluded from the image. The distribution of the hot gas is fairly symmetric. To illustrate the extent and distribution of the stellar light, the K-band intensity levels are overplotted. The contour levels are logarithmically scaled from $2\times10^7 \ \rm{L_{K,\odot} \ arcsec^{-2}}$ to $5\times10^8 \ \rm{L_{K,\odot} \ arcsec^{-2}}$. Within the outermost contour level, the enclosed light is $\sim$$3.2\times10^{11} \ \rm{L_{K,\odot}}$, which is $\sim$$76\%$ of the total K-band luminosity of NGC1961.}
\vspace{0.5cm}
     \label{fig:img_1961}
  \end{center}
\end{figure*}

\begin{figure*}[t]
  \begin{center}
    \leavevmode
      \epsfxsize=5in\epsfbox{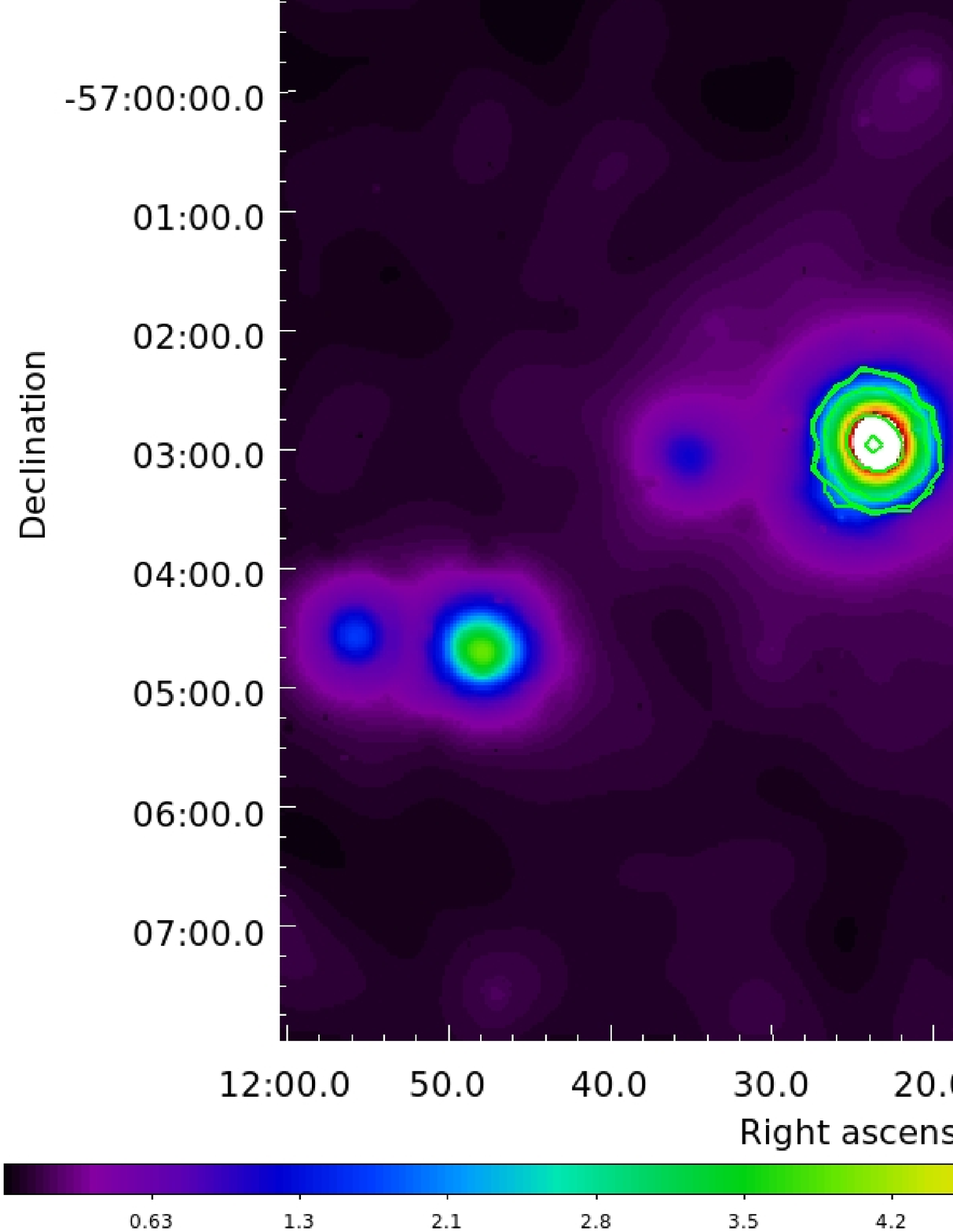}
      \caption{Denoised surface brightness image of NGC6753 based on the combined \textit{XMM-Newton} EPIC datasets. The side length of the image is $10\arcmin$, which corresponds to $126.7$ kpc at the distance of the galaxy. The diffuse X-ray emission, whose spectral signature is that of hot gas, extends beyond the optical radius of the galaxy. Note that point sources are not excluded from the image.  The overplotted K-band intensity levels demonstrate the distribution and extent of the stellar light. The contour levels are logarithmically scaled from $2\times10^7 \ \rm{L_{K,\odot} \ arcsec^{-2}}$ to $5\times10^8 \ \rm{L_{K,\odot} \ arcsec^{-2}}$. Within the outermost contour level the enclosed light is $\sim$$2.8\times10^{11} \ \rm{L_{K,\odot}}$, which is $\sim$$88\%$ of the total K-band lumiosity of the galaxy.}
\vspace{0.5cm}
     \label{fig:img_6753}
  \end{center}
\end{figure*}

\section{Data reduction}
\subsection{X-ray observations}
\label{sec:xmm}
NGC1961 was observed with the European Photon Imaging Camera (EPIC) aboard XMM-Newton in two observations (ObsID: 0673170101, 0673170301) for a total of 73.7 ks, whereas NGC6753 was observed in one pointing (ObsId: 0673170201) for 73.9 ks. Further details about the observations are listed in Table  \ref{tab:list2}.  

Since the main steps of the data reduction are identical with those outlined in  \citet{bourdin08}, here we only emphasize the major points. As a first step, we processed the event lists using the XMM Science Analysis System (SAS) version 10.0 and Current Calibration Files (CCF). We further identified and removed the flare contaminated time periods for each observation and camera separately, employing a two-step filtering of the high energy ($10-12$ keV) and softer energy events ($1-5$ keV). The clean exposure times are $30 - 80\%$ of the original exposures. To increase the signal-to-noise ratio, we combined the data
from the EPIC MOS and EPIC PN instruments, and also merged the two pointings of NGC1961. Since the goal of our study is to characterize the diffuse emission, bright point sources were identified using the combined EPIC PN and MOS data. To do so, we selected the PSF smeared point sources up to a radius where their surface brightness reaches the background noise fluctuation. These radii are derived from the analysis of wavelet denoised images of the field of view using the Source Extractor software \citep{bertin96}.  With typical radii in the range of $10-30\arcsec$, the detected point sources are masked out in the further analysis of the diffuse emission.

\begin{figure*}[t]
  \begin{center}
    \leavevmode
      \epsfxsize=8.7cm\epsfbox{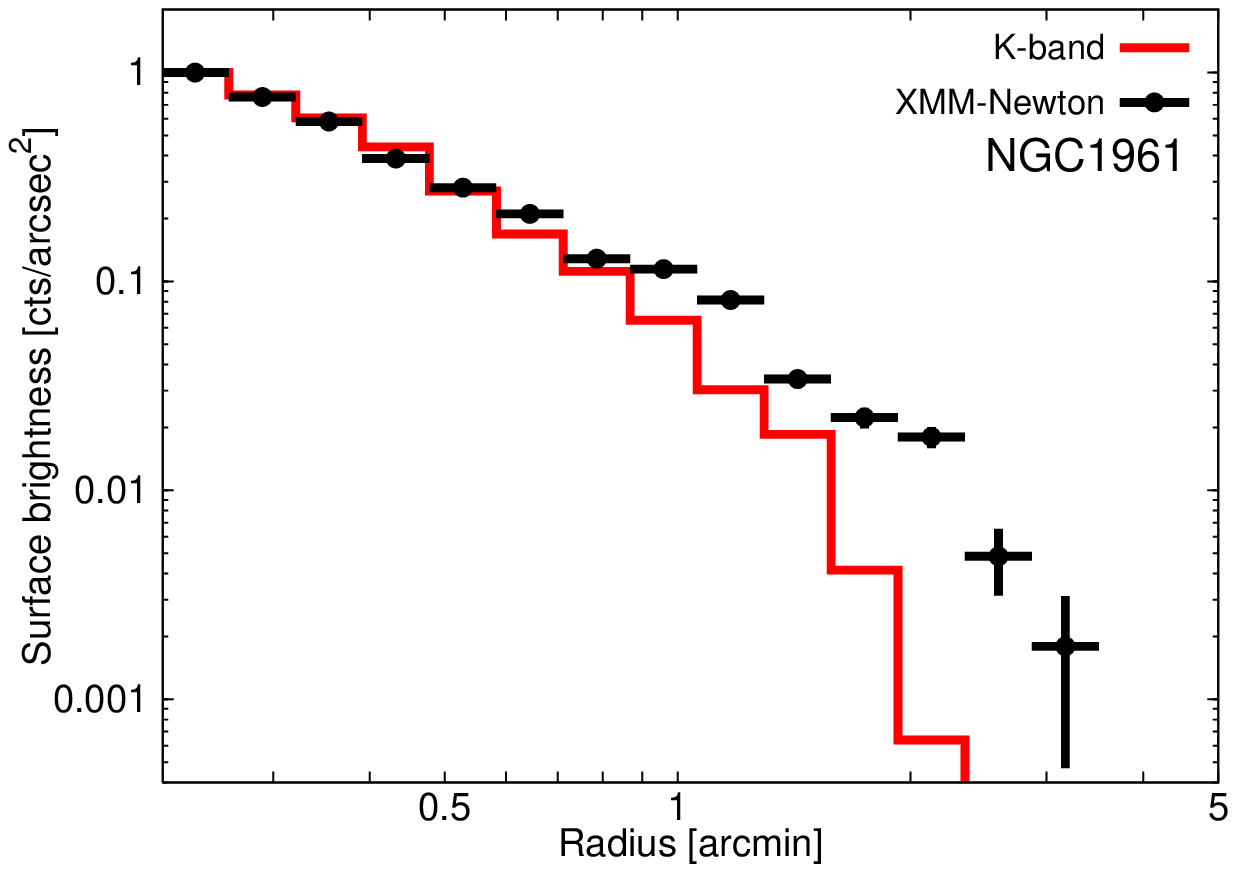}
\hspace{0.2cm} 
      \epsfxsize=8.7cm\epsfbox{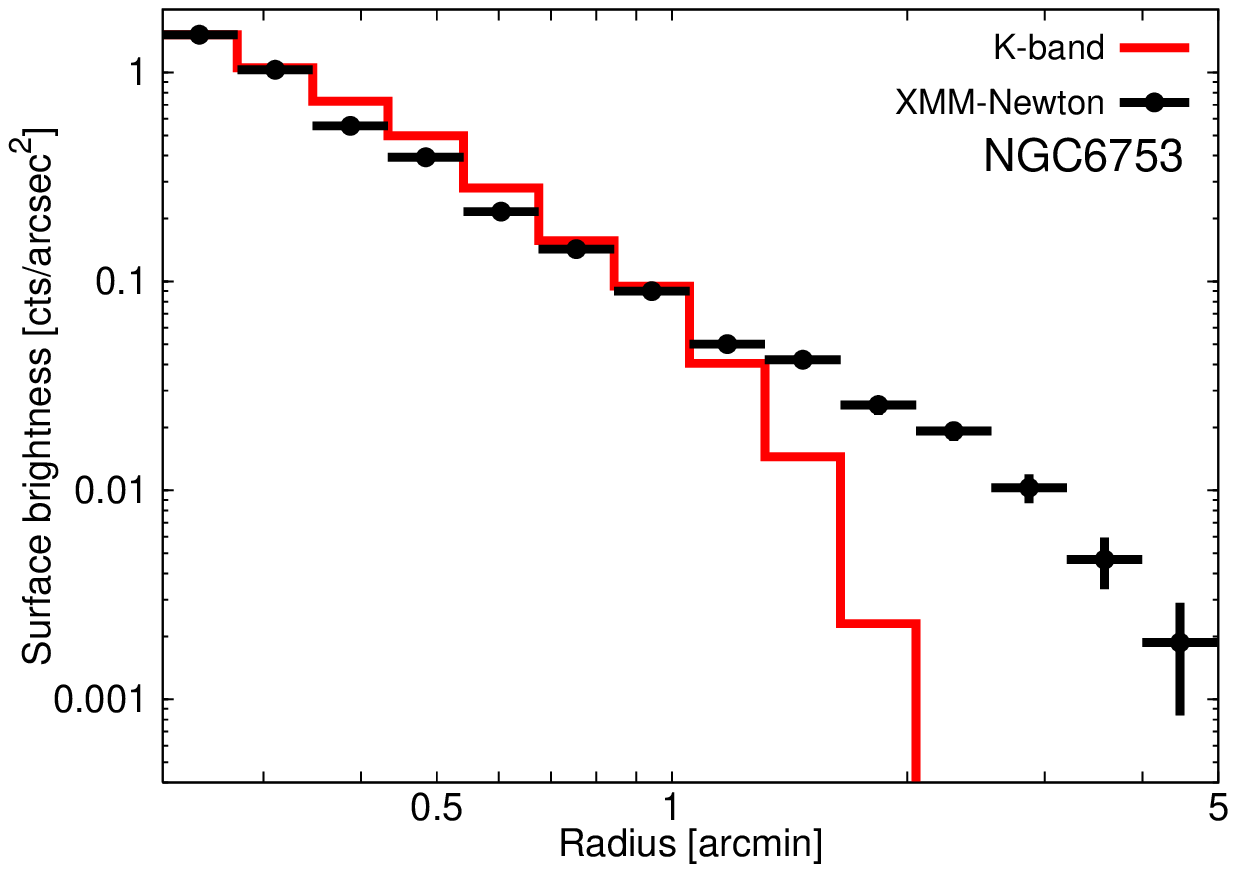}
      \caption{Surface brightness distribution of the $0.3-2 $ keV band diffuse X-ray emission for NGC1961 (left panel) and NGC6753 (right panel) obtained from the combined data of \textit{XMM-Newton} EPIC PN and EPIC MOS cameras. To construct the profiles, we used circular annuli centered on the optical centroid of each galaxy. Vignetting correction is applied and all background components are subtracted.  The X-ray light distributions are compared with the K-band light profiles. The K-band profiles are   normalized to match the level of X-ray emission in the innermost bins.}
\vspace{0.5cm}
     \label{fig:profile}
  \end{center}
\end{figure*}

\begin{figure*}
  \begin{center}
    \leavevmode
      \epsfxsize=8.7cm\epsfbox{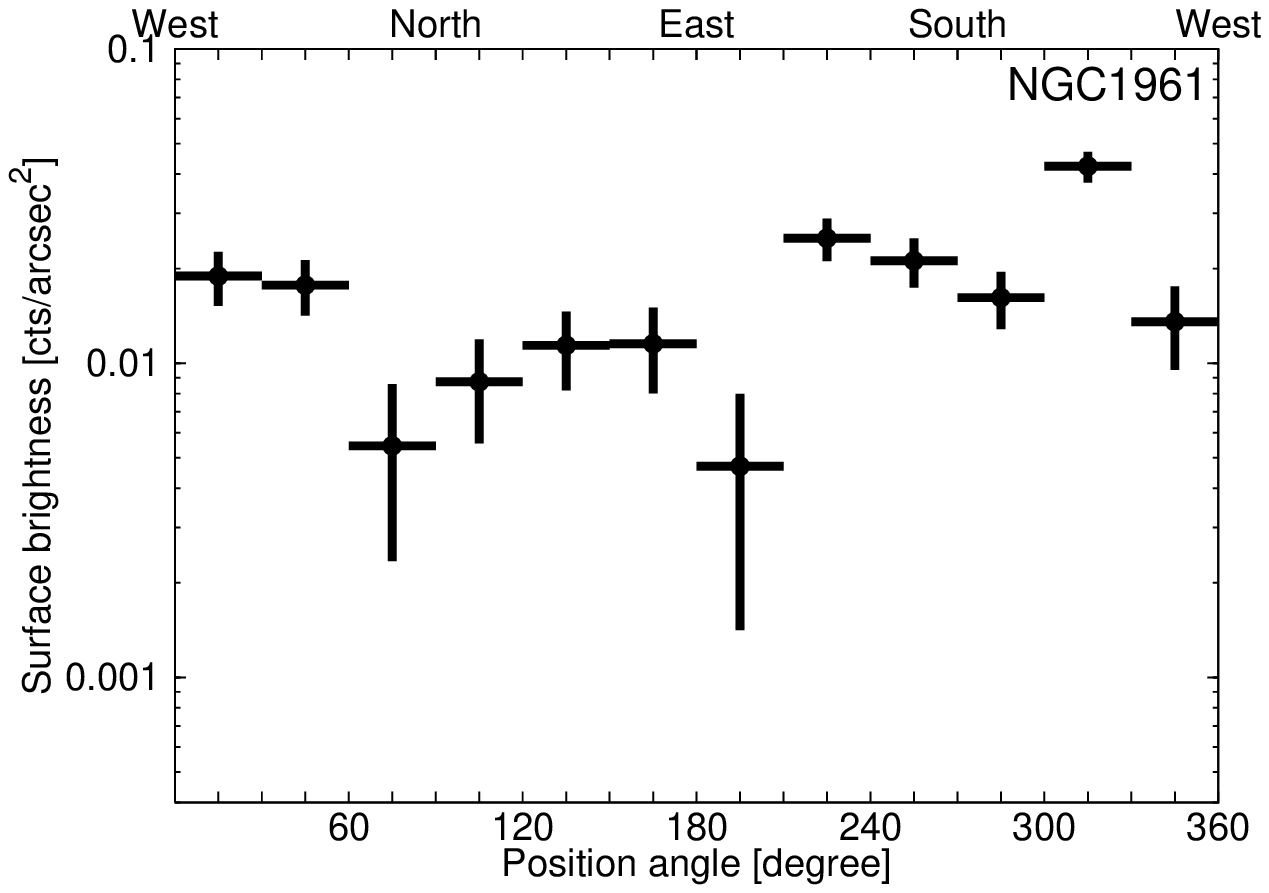}
\hspace{0.2cm} 
      \epsfxsize=8.7cm\epsfbox{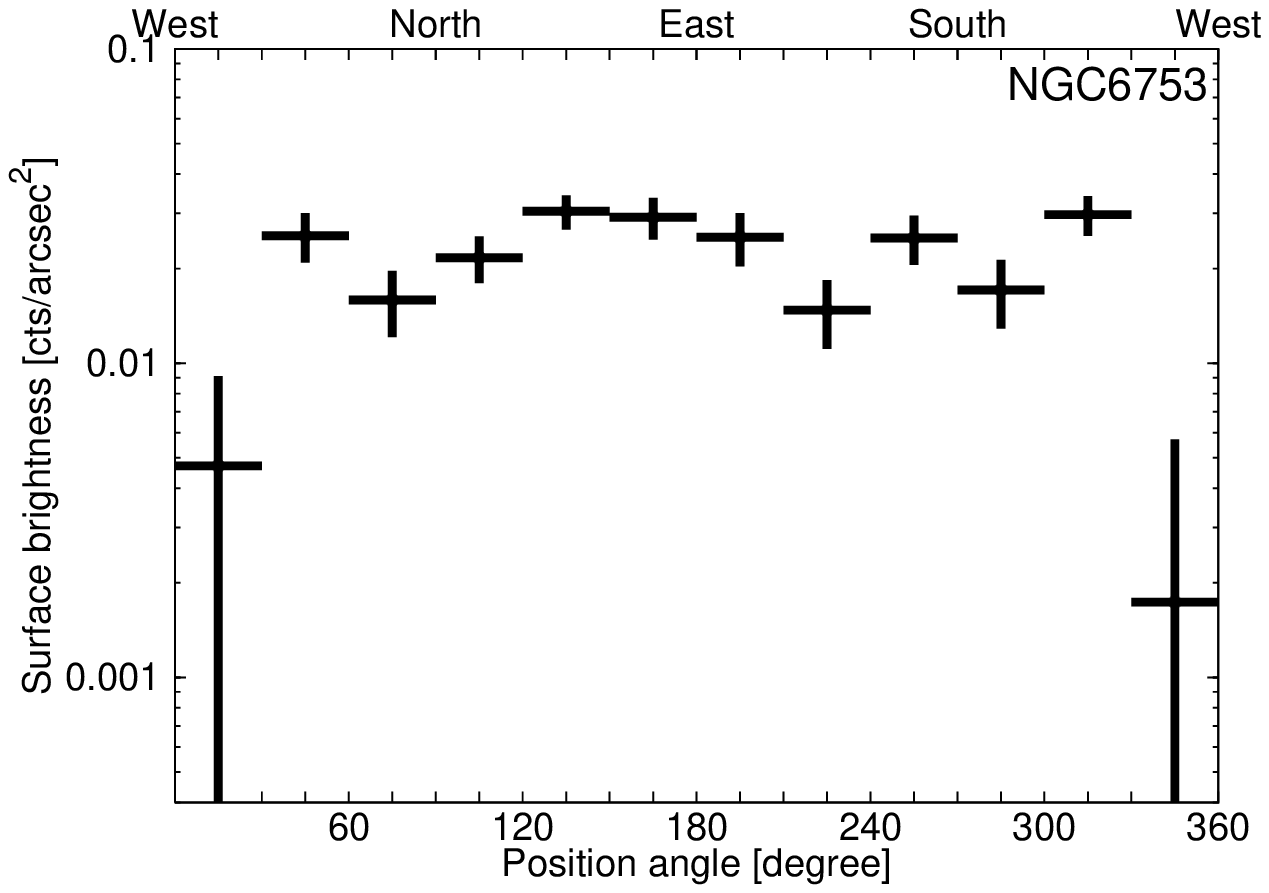}
      \caption{Surface brightness distribution of the $0.3-2 $ keV band diffuse X-ray emission for NGC1961 (left panel) and NGC6753 (right panel) obtained from the combined data of \textit{XMM-Newton} EPIC PN and EPIC MOS cameras. The profiles are obtained from circular wedges, whose inner and outer radii are $1.25\arcmin-3 \arcmin$ for both galaxies. Vignetting correction is applied and the background components are subtracted. As shown in the upper x-axis, $90\degr$ corresponds to north and $180\degr$ corresponds to east.}
\vspace{0.5cm}
     \label{fig:wedges}
  \end{center}
\end{figure*}

Images and surface brightness profiles of the diffuse emission are
presented in Figures \ref{fig:img_1961}, \ref{fig:img_6753},  \ref{fig:profile}, and  \ref{fig:wedges}. All of these are corrected for effective areas and background noise models. Spatially and spectrally sampled, the background model consists of (i) the cosmic-ray induced particle background, (ii) the expected out-of-time count rate, (iii) a residual emission associated with soft protons (iv) a vignetted, extragalactic power-law component, (v) a vignetted two-phase thermal plasma model designed to account for emission from the Milky Way. Derived from ``closed-filter'' data sets, the cosmic-ray induced particle background
includes spatially variable continua and fluorescence lines for each detector CCD, as detailed in \citet{bourdin13}. The out-of-time count rate is fixed in each energy band to $6.3\%$ of all photon counts registered along the CCD columns. A soft proton excess was detected and modelled as a power-low in the case of NGC1961. The extragalactic power-law index is fixed to 1.42 \citep[see e.g.][]{lumb02}, while the Galactic foregrounds are modelled by the sum of two thermal bremstrahlung spectra with Solar metal abundances ($kT_1 = 0.099$ keV and $kT_2 = 0.248$ keV), accounting for the Galactic transabsorption emission \citep{kuntz00}. All of these components were first normalized using a spectral fit to the source data in an annulus with $10\arcmin-15\arcmin$ radii, and then individually rescaled within each image pixel and energy bin using the appropriate vignetting factors.  The accuracy of the model can be seen in Figure \ref{fig:spec}.

Since accurate background subtraction is crucial in our study, we checked the accuracy of the method discussed above by two means. First, we accounted for the background components using the EPIC ``blank sky'' files \citep{carter07}. These background files were tailor made for our observations, that is they were selected to approximately match the  galactic foreground absorption values of NGC1961 and NGC6753 and the revolution number of the\textit{XMM-Newton} observations.  To match the observed count rates, we re-normalized the background files using the $11-12$ keV band count rates, where the effective area of \textit{XMM-Newton} is negligible. Second, we analyzed the publicly available \textit{Chandra} observations of NGC1961 (ObsId: 10528-10533). To reduce the data and subtract the background components, we followed the method described in \citet{bogdan08}. Using these two different methods, we constructed X-ray surface brightness distributions and extracted X-ray energy spectra of the studied galaxies (only NGC1961 for \textit{Chandra}). These were compared with the profiles and spectra obtained from our standard background subtraction method, which models all background components. The resulting profiles agree with each other within statistical uncertainties at all radii, moreover the obtained best-fit parameters of the spectra also agree within the $1\sigma$ uncertainties. Since all three background subtraction methods result in comparable profiles and spectra, it implies that our background subtraction method is robust and accurate.

\subsection{Near-infrared observations}
\label{sec:2mass}
To trace the stellar light of NGC1961 and NGC6753, we rely on the K-band data of the Two-Micron All Sky Survey (2MASS) Large Galaxy Atlas \citep{jarrett03}. The observed background subtracted K-band counts ($S$) are converted to physical units using
\begin{eqnarray}
m_{\rm{K}}= \rm{KMAGZP} - 2.5 \log S \ , 
\end{eqnarray}
where $m_{\rm{K}}$ is the apparent K-band magnitude, and $\rm{KMAGZP}$ is the zero point magnitude given in the image header. The derived $m_{\rm{K}}$ is converted to an absolute magnitude and luminosity assuming that the absolute K-band magnitude of the Sun is $ M_{K,\odot} = 3.28 $ mag. The total K-band luminosities of NGC1961 and NGC6753 are $5.4\times10^{11} \ \rm{L_{K,\odot}}$ and $3.9\times10^{11} \ \rm{L_{K,\odot}}$, respectively. 

We compute the total stellar mass of NGC1961 and NGC6753 from their K-band luminosities and from the K-band mass-to-light ratios (Table \ref{tab:list1}). The mass-to-light ratios are derived from the $B-V$ color indices and results of galaxy evolution modeling \citep{bell03}. The mass-to-light ratios are $0.78 \ \rm{M_{\odot}/L_{K,\odot}}$ for NGC1961 and $0.81 \ \rm{M_{\odot}/L_{K,\odot}}$ for NGC6753, hence their total stellar masses  are $4.2\times10^{11} \ \rm{M_{\odot}}$ and $3.2\times10^{11} \ \rm{M_{\odot}}$, respectively. 

While the uncertainties associated with the K-band photometry are $\lesssim3\%$, the mass-to-light ratios obtained from the $B-V$ color indices and the \citet{bell03} relation employ a number of assumptions and are affected by notable systematic uncertainties. The most important assumtion is the applied stellar initial mass function, while the systematic uncertainties are dominated by the effects of dust and starbursts, which may result in $\sim$$25\%$ uncertainty. The importance and significance of various assumptions and uncertainties is discussed in full particulars in \citet{bell03}.

\subsection{Far-infrared observations}
\label{sec:fir}
We compute the star-formation rates (SFRs)  of NGC1961 and NGC6753 based on their far-infrared emission observed by the \textit{Infrared Astronomical Satellite} (\textit{IRAS}), and applying the relation described by \citet{kennicutt98}:
\begin{eqnarray}
\rm{SFR_{IR} \ (M_{\odot} \ yr^{-1})} = 4.5\times10^{-44}  L_{\rm{IR}}  \ (erg \ s^{-1}) \ ,
\end{eqnarray}
where $ L_{\rm{IR}}$ is the total infrared luminosity in the $8-1000 \ \rm{\mu m}$ regime.  To obtain the total infrared luminosity, we follow the procedure of \citet{meurer99}. The total far-infrared flux ($F_{\rm{FIR}}$) in the $40-120 \ \rm{\mu m}$ range \citep{helou88} is 
\begin{eqnarray}
F_{\rm{FIR}} = 1.26 \times 10^{-11}  (2.58 F_{\rm{60\mu m}} + F_{\rm{100\mu m}}) \ ,
\end{eqnarray}
where $F_{\rm{60\mu m}}$ and $F_{\rm{100\mu m}}$ are the \textit{IRAS} fluxes at $60 \ \rm{\mu m}$ and $100 \ \rm{\mu m}$ \citep{sanders03}. To convert the flux to a luminosity, we use $L_{\rm{FIR}} = 4 \pi D^2_L F_{\rm{FIR}}$, where $D^2_L$ is the luminosity distance of the galaxy. Since the \citet{kennicutt98} relation refers to the total infrared luminosity, we convert $L_{\rm{FIR}} $ to $L_{\rm{IR}}$ applying the formula of \citet{calzetti00}:
\begin{eqnarray}
L_{\rm{IR}} \sim 1.75 L_{\rm{FIR}} \ . 
\end{eqnarray}
Although this conversion refers to the  $1-1000 \ \rm{\mu m}$ band and not to the  $8-1000 \ \rm{\mu m}$ band, the contribution from the  $1-8 \ \rm{\mu m}$ range is not expected to exceed the few per cent level \citep{calzetti00}, yielding slightly  overestimated SFRs. We thus obtain SFRs of $15.5 \ \rm{M_{\odot} \ yr^{-1}}$ in NGC1961 and $11.8 \ \rm{M_{\odot} \ yr^{-1}}$ in NGC6753, implying moderate ongoing star-formation.

\begin{figure*}[t]
  \begin{center}
    \leavevmode
      \epsfxsize=8.5cm\epsfbox{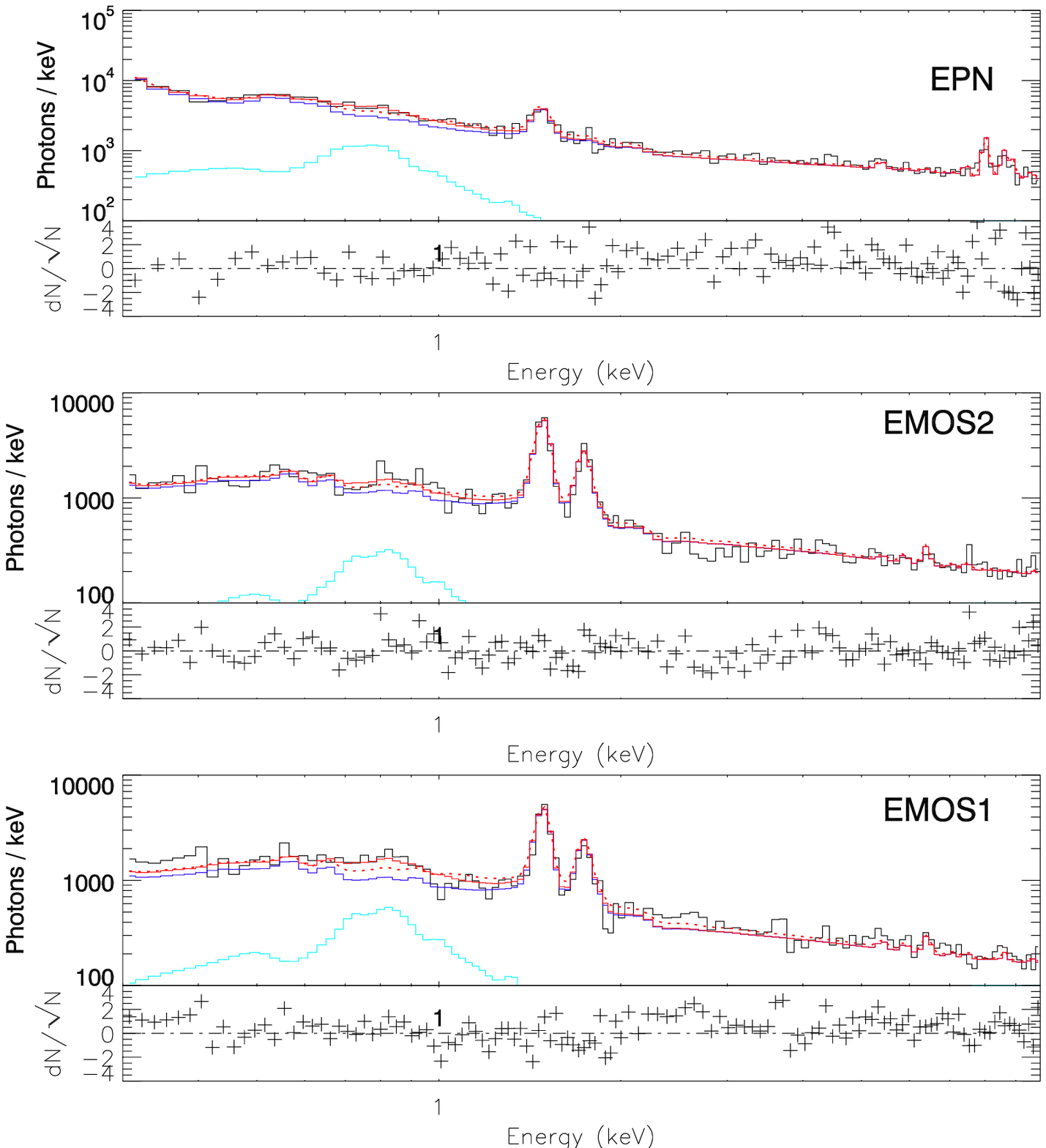}
\hspace{0.2cm} 
      \epsfxsize=8.5cm\epsfbox{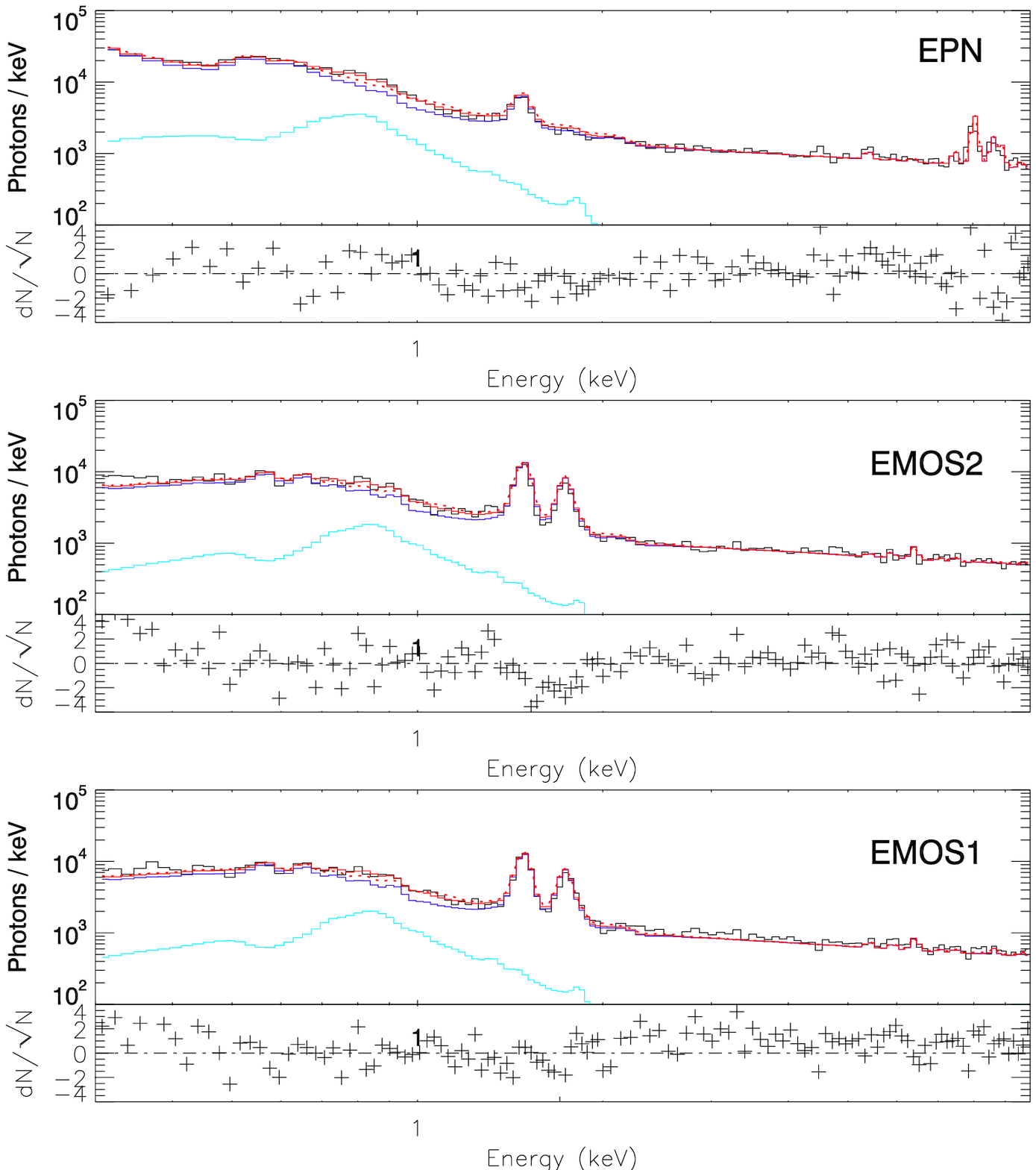}
      \caption{X-ray energy spectra of NGC1961 (left panel) and NGC6753 (right panel) in the $0.3-10$ keV band obtained from the \textit{XMM-Newton}  EPIC PN (top panel) MOS1 (middle) and MOS2 (bottom) cameras. The depicted spectra are extracted for the $(0.05-0.15)r_{200}$ radial range. The contribution of bright point sources are removed. The spectra are fit with an absorbed thermal plasma emission model. The black lines show the data set and the red lines represent the overall fit. Dark blue shows the overall background components, whereas light blue is the emission associated with the thermal spectrum. The bottom panels of each plot show the  residuals with respect to the best-fit thermal model.}
\vspace{0.5cm}
     \label{fig:spec}
  \end{center}
\end{figure*}

\section{Results}
To detect and characterize the physical properties of the hot X-ray coronae around NGC1961 and NGC6753, we employ three different techniques. First, we produce surface brightness images in the $0.3-2$ keV band to map the spatial distribution of the diffuse X-ray emission around the galaxies. Second, we build X-ray surface brightness profiles, which we compare with the stellar light distrubiton. Third, to measure the physical properties of the coronal  X-ray emisson, we extract the X-ray energy spectra. 

\subsection{Images}
\label{sec:images}
To map the large-scale spatial distribution of the diffuse emission around NGC1961 and NGC6753, we construct surface brightness images. The images are produced in the $0.3-2$ keV energy band. This  particular choice of the energy range is advantageous because 1)  the hot X-ray gas is predicted to have sub-keV temperatures (Section \ref{sec:temperature}), hence the bulk of its emission is produced in the soft band, 2) the effective area of \textit{XMM-Newton} detectors is the highest, and 3) the level of instrumental and cosmic X-ray background is fairly low in the $0.3-2$ keV band. 

In Figures \ref{fig:ngc1961} and \ref{fig:ngc6753} we show the unprocessed $0.3-2$ keV band EPIC MOS images of the central $10\arcmin\times10\arcmin$ regions of NGC1961 and NGC6753, respectively. The X-ray images are compared with optical images from the Digitized Sky Survey (DSS). An apparent X-ray glow is associated with both galaxies within  their optical extent. As discussed in Section \ref{sec:introduction}, this X-ray emission originates from the population of resolved and unresolved discrete sources, while truly diffuse gaseous emission also contributes. However, in the unprocessed images, the diffuse X-ray emission cannot be identified beyond the optical extent of the galaxies,  due to the faint nature of the hot X-ray coronae  and the relatively high background level. 

In order to better visualize  the faint X-ray coronae of the sample galaxies beyond their optical extent, we produce ``denoised'' surface brightness maps of the photon images following \citet{starck09}. The images are background subtracted and vignetting corrected. To facilitate the comparison with the stellar light distribution, logarithmically-scaled contour levels of the K-band light are overplotted.

The ``denoised'' $0.3-2$ keV band X-ray images suggest the presence of extended diffuse emission  around both NGC1961 (Figure \ref{fig:img_1961}) and NGC6753 (Figure \ref{fig:img_6753}), whose extents significantly exceed those of the stellar light. Whereas the dominant fraction of the optical light is confined to within $\sim$$1.5\arcmin$, the diffuse X-ray emission appears to extend to several arcminutes in both galaxies. As discussed in Section \ref{sec:spectra}, the spectral signature of this emission is that of hot gas.  The morphology of the X-ray emission is fairly symmetric, thereby hinting that the hot gas resides in hydrostatic equilibrium in the dark matter halos of NGC1961 and NGC6753. The characteristics of these X-ray coronae are further discussed in Sections \ref{sec:profiles} and \ref{sec:spectra}.

\subsection{Profiles}
\label{sec:profiles}
To further study the spatial distribution and  the morphology of the diffuse X-ray emission around NGC1961 and NGC6753, we construct X-ray surface brightness profiles. The profiles are extracted from the $0.3-2$ keV energy band using the combined \textit{XMM-Newton} EPIC data. From the profiles all background components are subtracted, vignetting correction is applied, and bright resolved point sources are masked out. 

To extract the X-ray surface brightness profiles, we employ circular annuli centered on the optical center of each galaxy. The distribution of the X-ray light is compared with the stellar light distribution, which is measured with the 2MASS K-band images. The K-band light profiles are extracted using the same circular annuli, moreover the locations of X-ray sources are masked out from the K-band images, along with the population of bright foreground stars. Note that the uncertaintes in the X-ray surface brightness are of a statistical nature, but at large offsets the systematic uncertainties dominate.  The profiles, depicted in Figure \ref{fig:profile}, confirm that the diffuse X-ray emission extends well beyond the optical radii for NGC1961 and NGC6753. Whereas the stellar light is confined to within $20-25$ kpc ($\sim$$1.5\arcmin$), the diffuse X-ray emission is detected out to $\sim$$57$ kpc ($3.5\arcmin$) in NGC1961 and $\sim$$63$ kpc ($5\arcmin$) in NGC6753. This implies that the diffuse X-ray emission at large radii cannot be associated with the stellar body of the host galaxies. Interestingly, the observed surface brightness distribution of NGC6753 is flatter than that obtained for NGC1961, implying that the emission may extend beyond $\sim5\arcmin$. However, beyond this radius the surface brightness drops rapidly and -- due to the rather high instrumental background level of \textit{XMM-Newton} -- statistically significant emission cannot be detected. To explore the hot X-ray corona of NGC6753 beyond $\sim5\arcmin$, deep \textit{Chandra} observations are desirable. 

In the ``denoised'' images, the diffuse emission appears to be fairly symmetric. To further probe the spatial distribution of the X-ray emission, we produce X-ray surface brightness profiles using circular wedges centered on the centroid of each galaxy. Since we aim to probe the surface brightness beyond the optical extent, the wedges are extracted from annuli with radii of $1.25\arcmin-3\arcmin$ for both galaxies. Since the stellar light is negligible in this region, we do not plot the  level of renormalized K-band light. As before, the uncertainties are of a purely statistical nature. The circular profiles,  shown in Figure \ref{fig:wedges}, demonstrate that the hot X-ray gas has a fairly uniform distribution around both galaxies at every position angle. The spatial distributions are morphologically consistent with warm gas residing in hydrostatic equilibrium in the potential well of the galaxies, and are morphologically inconsistent with a starburst or AGN driven bipolar outflow.

\subsection{Spectra}
\label{sec:spectra}
\subsubsection{Observed parameters in the $(0.05-0.15)r_{200}$ region}
To study the nature and physical properties of the diffuse X-ray emission around NGC1961 and NGC6753, we extract X-ray energy spectra. As we aim to probe the hot gas beyond the optical radii, we exclude the central regions of the galaxies. To facilitate the comparison with structure formation simulations, we convert all radial distances to $r_{200}$ units (Section \ref{sec:galaxies}). For both galaxies, we extract spectra from circular annuli with $(0.05-0.15)r_{200}$, which correspond to the physical distances of $23.5-70.5$ kpc for NGC1961 and $22.0-66.0$ kpc for NGC6753. 

In Figure \ref{fig:spec} we show the EPIC PN and EPIC MOS spectra for NGC1961 and NGC6753, which demonstrate the presence of a soft X-ray emitting component at energies below $2$ keV. Within the studied region we detect $1778$ and $4858$ net counts from NGC1961 and NGC6753. Defining the signal-to-noise ratio as $\mathrm{SNR}= (C_S/C_B^{0.5})$, where $C_s$ is the number of net counts and $C_B$ is the number of background counts, we obtain $\mathrm{SNR_{NGC1961}}=13.1$ and  $\mathrm{SNR_{NGC6753}}=22.3$ for NGC1961 and NGC6753, respectively. To describe the soft X-ray emitting component, we employ an absorbed thermal plasma model (\textsc{wabs*apec} in \textsc{Xspec}). When fitting the spectra, the column density is fixed  at the Galactic value \citep{dickey90} and the abundance is left free to vary. For  spectral fitting the abundance table of \citet{grevesse98} is used.  The thermal components describe the spectra with reduced $\chi^2=1.27$ for NGC1961 and reduced $\chi^2=1.79$ for NGC6753. These spectra indicate that the diffuse X-ray emission originates from hot gas. However, these fits are formally not acceptable, which may be the result of the complex thermal and/or abundance structure of the hot gas or it could be an artifact caused by the insufficient energy resolution of the detectors.

Based on the best-fit spectra we measure the average parameters of the hot gas within the $(0.05-0.15)r_{200}$ region. For both NGC1961 and NGC6753, we obtained similar gas temperatures ($kT\sim0.6-0.7$ keV) and sub-solar abundances ($0.12-0.13$ Solar). From the emission measure of the spectra ($\int n_e n_H dV$), we compute the average electron number density, the confined gas mass, and the average cooling time. As a caveat we mention that the emission measure and abundances of the thermal component are strongly anticorrelated (Section \ref{sec:ironbias}). To compute the hot gas mass, we assumed a spherically symmetric gas distribution, constant density, and a volume filling factor of unity. Given the volume of the studied regions, we obtained hot gas masses of about $\sim$$10^{10}  \ \rm{M_{\odot}}$ and electron number densities of  $(3.5-4)\times10^{-4} \ \rm{cm^{-3}}$ for both galaxies. The cooling time of the gas is computed from $t_{\rm{cool}}= (3kT)/(n_e  \Lambda(T))$, where $\Lambda(T)$ is the cooling function at the best-fit temperature and abundance. For both NGC1961 and NGC6753 we obtain $t_{\rm{cool}}\sim50$ Gyrs, much longer than the Hubble-time. The uncertainties of the emission measure (and other quantities derived from its value)  are estimated from the parameter covariance matrix by assuming a quadratic $\chi^2$ distribution.  The   parameters of the warm gas are listed in Table \ref{tab:values}.

\begin{table*}
\caption{The parameters of the extended hot gaseous coronae around NGC1961 and NGC6753.}
\begin{minipage}{18cm}
\renewcommand{\arraystretch}{1.3}
\centering
\begin{tabular}{c c c c c c c c}
\hline 
Galaxy&  Radial range &$ \int n_e n_H dV $ & $kT$ &Abundance &$M_{\rm{gas}}$ & $n_{\rm{e}}$  & $t_{\rm{cool}}$ \\ 
 & $r_{200}$ &$ \rm{cm^{-3}}$ & keV & & $M_{\rm{\odot}}$ & $\rm{cm^{-3}}$ & Gyr  \\ 
\hline
NGC1961 & $0.05-0.15$ & $(5.2\pm1.5)\times10^{63}$ & $0.61^{+0.10}_{-0.13}$ & $0.12\pm0.03$ & $(1.2\pm0.2)\times10^{10}$ & $(3.5\pm0.5)\times10^{-4}$ & $45\pm6$ \\
NGC6753 & $0.05-0.15$ &  $(5.3\pm0.8)\times10^{63}$ &$0.69^{+0.06}_{-0.07}$ & $0.13\pm0.02$ & $(1.1\pm0.1)\times10^{10}$ &$(3.9\pm0.3)\times10^{-4}$ & $51\pm4$ \\
\hline
NGC1961 & $0.15-0.30$ & $<1.2\times10^{63}$ & $0.6^{\dagger}$ & $0.12^{\dagger}$ & $<1.6\times10^{10}$ & $<6.4\times10^{-5}$ & $>258$ \\
NGC6753 & $0.15-0.30$ &  $<9.6\times10^{62}$ &$0.6^{\dagger}$ & $0.13^{\dagger}$ & $<1.3\times10^{10}$ & $<6.2\times10^{-5}$ & $>266$ \\

\hline \\
\end{tabular} 
\end{minipage}
$^{\dagger}$ These parameters were fixed at the given value.

\label{tab:values}
\end{table*}

\begin{table*}
\caption{The flux and luminosity of the extended hot gaseous coronae  around NGC1961 and NGC6753.}
\begin{minipage}{18cm}
\renewcommand{\arraystretch}{1.3}
\centering
\begin{tabular}{c c c c c c}
\hline 
Galaxy& Radial range & $F_{\rm{0.5-2keV,abs}}$& $L_{\rm{0.5-2keV,abs}}$&$L_{\rm{0.5-2keV,unabs}}$  &$L_{\rm{bol}}$ \\ 
 &  $r_{200}$ &$ \rm{erg \ s^{-1} \ cm^{-2}}$ & $ \rm{erg \ s^{-1}}$ & $\rm{erg \ s^{-1}}$ & $\rm{erg \ s^{-1}}$   \\ 
\hline
NGC1961 & $0.05-0.15$ & $(5.4\pm1.6)\times10^{-14}$ &  $(2.0\pm0.6)\times10^{40}$ & $(2.8\pm0.8)\times10^{40}$ & $(5.8\pm1.7)\times10^{40}$ \\
NGC6753 & $0.05-0.15$ & $(1.1\pm0.2)\times10^{-13}$ &$(2.5\pm0.4)\times10^{40}$ & $(3.1\pm0.5)\times10^{40}$ & $(6.3\pm0.9)\times10^{40}$ \\
\hline 
NGC1961 & $0.15-0.30$ & $<1.3\times10^{-14}$ &$<4.8\times10^{39}$ & $<6.7\times10^{39}$ & $<1.4\times10^{40}$ \\
NGC6753 & $0.15-0.30$ & $<1.1\times10^{-14}$ &$<2.5\times10^{39}$ & $<3.1\times10^{39}$ & $<6.5\times10^{39}$ \\

\hline \\
\end{tabular} 
\end{minipage}

\label{tab:luminosity}
\end{table*}

From the best-fit spectra, we also compute the X-ray luminosity of the hot gas in the $(0.05-0.15)r_{200}$ radial bin. In the $0.5-2$ keV band, we obtain  absorbed luminosities of $L_{\rm{0.5-2keV,abs}} =(2.0\pm0.6)\times10^{40} \ \rm{erg \ s^{-1}}$ for NGC1961 and $L_{\rm{0.5-2keV,abs}} =(2.5\pm0.4)\times10^{40} \ \rm{erg \ s^{-1}}$. Using the Galactic column densities \citep{dickey90}, we convert these values to unabsorbed luminosities. Finally, we also derive the corresponding bolometric luminosities of the X-ray coronae, and we find $L_{\rm{bol}} = (5.8\pm1.7)\times10^{40} \ \rm{erg \ s^{-1}}$ and $L_{\rm{bol}} = (6.3\pm0.9)\times10^{40} \ \rm{erg \ s^{-1}}$ for NGC1961 and NGC6753, respectively. The obtained fluxes and luminosities are listed in  Table \ref{tab:luminosity}.

\subsubsection{Upper limits in the $(0.15-0.30)r_{200}$ region}
In the $(0.15-0.30)r_{200}$ region, we do not detect a statistically significant X-ray emission above the background level in either NGC1961 or NGC6753. In the absence of hot gas detection beyond $\sim$$0.15r_{200}$, we place upper limits on its physical parameters in the $(0.15-0.30)r_{200}$ region. This region is well within the \textit{XMM-Newton} FOV, and corresponds to the radial range of $70.5-141.0$ kpc for NGC1961 and $66.0-132.0$ kpc for NGC6753. From the observed count rates within these regions, we derive $3\sigma$ upper limits using the EPIC PN data. To constrain the gas properties, we rely on the gas parameters observed within the $(0.05-0.15)r_{200}$ region. In particular, we assume the abundance of $0.12$ Solar for NGC1961 and $0.13$ Solar for NGC6753, and we adopt  $kT=0.6$ keV for both galaxies. 

Using the appropriate response files, we derive upper limits on the emission measure, which we use to place upper limits on the confined gas mass, the electron number density, and the cooling time. Within the $(0.15-0.30)r_{200}$ region, the $3\sigma$ upper limit on the gas mass is $<1.6\times10^{10}  \ \rm{M_{\odot}}$ and $<1.3\times10^{10}  \ \rm{M_{\odot}}$ for NGC1961 and NGC6753, respectively. For the average electron number density, we derive  $<6.4\times10^{-5} \ \rm{cm^{-3}}$ in NGC1961 and $<6.2\times10^{-5} \ \rm{cm^{-3}}$  in NGC6753. The cooling times for both galaxies are longer than $250$ Gyrs, hence significantly exceed the Hubble-time, implying the quasi-static nature of the hot gas. The obtained upper limits are listed in Table \ref{tab:values}. 

Based on the upper limits on the emission measure, we also constrain the X-ray luminosities within the $(0.15-0.30)r_{200}$ region  (Table \ref{tab:luminosity}). The upper limits on the absorbed $0.5-2$ keV band luminosities are $<4.8\times10^{39} \ \rm{erg \ s^{-1}}$ and $< 2.5\times10^{39} \ \rm{erg \ s^{-1}}$ for NGC1961 and NGC6753, respectively.  These values correspond to the bolometric luminosities of $< 1.4\times10^{40} \ \rm{erg \ s^{-1}}$ in NGC1961 and $< 6.5\times10^{39} \ \rm{erg \ s^{-1}}$ in NGC6753.

\section{Comparison with numerical simulations}
\label{sec:simulation}
\subsection{Numerical codes}
\label{sec:codes}
In this paper the observed properties of X-ray coronae around NGC1961 and NGC6753 are confronted with hydrodynamical simulations of structure formation. In particular, we probe the recently developed moving mesh code \textsc{arepo} \citep{springel10} and the traditional SPH-based code \textsc{gadget}  \citep{springel05}. The codes and the methodology of the simulations are discussed  in full particulars in \citet{vogelsberger12}.

\begin{figure*}[!t]
  \begin{center}
    \leavevmode
      \epsfxsize=9cm\epsfbox{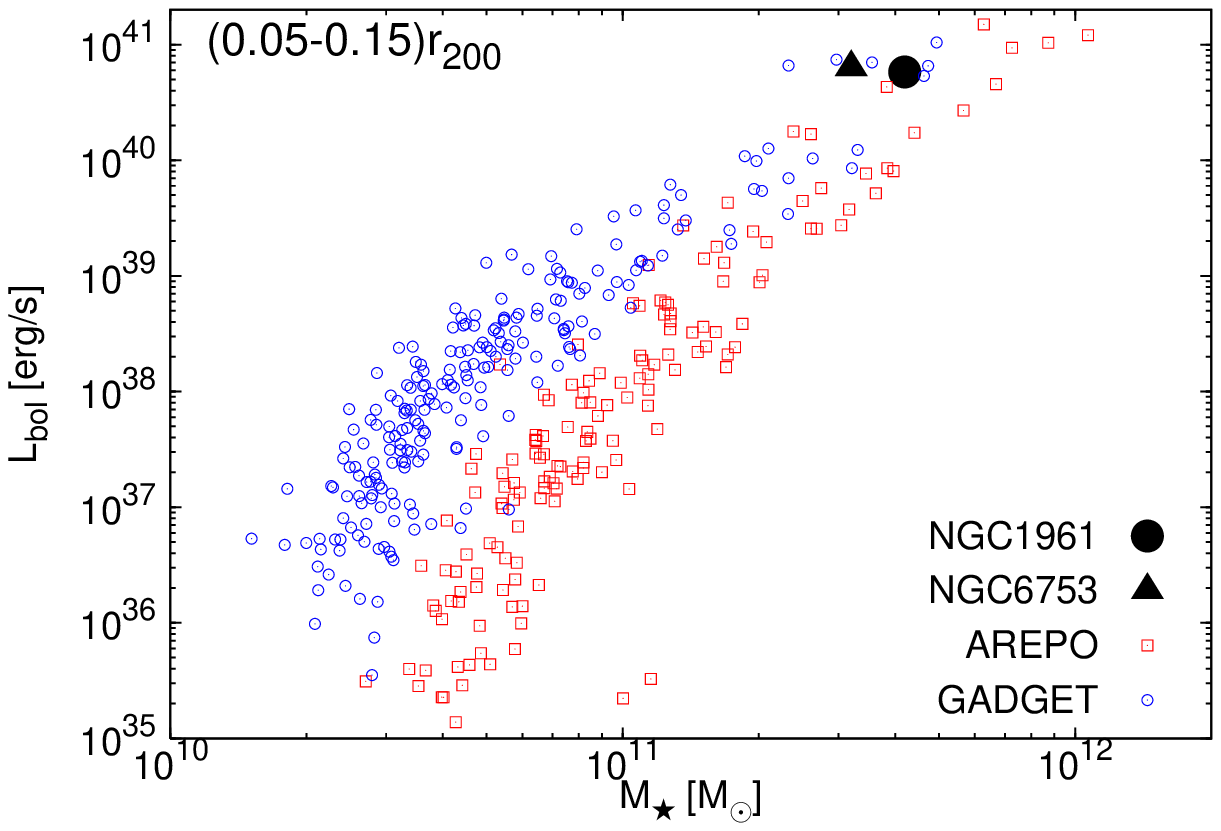}
\hspace{-0.1cm}
      \epsfxsize=8.6cm\epsfbox{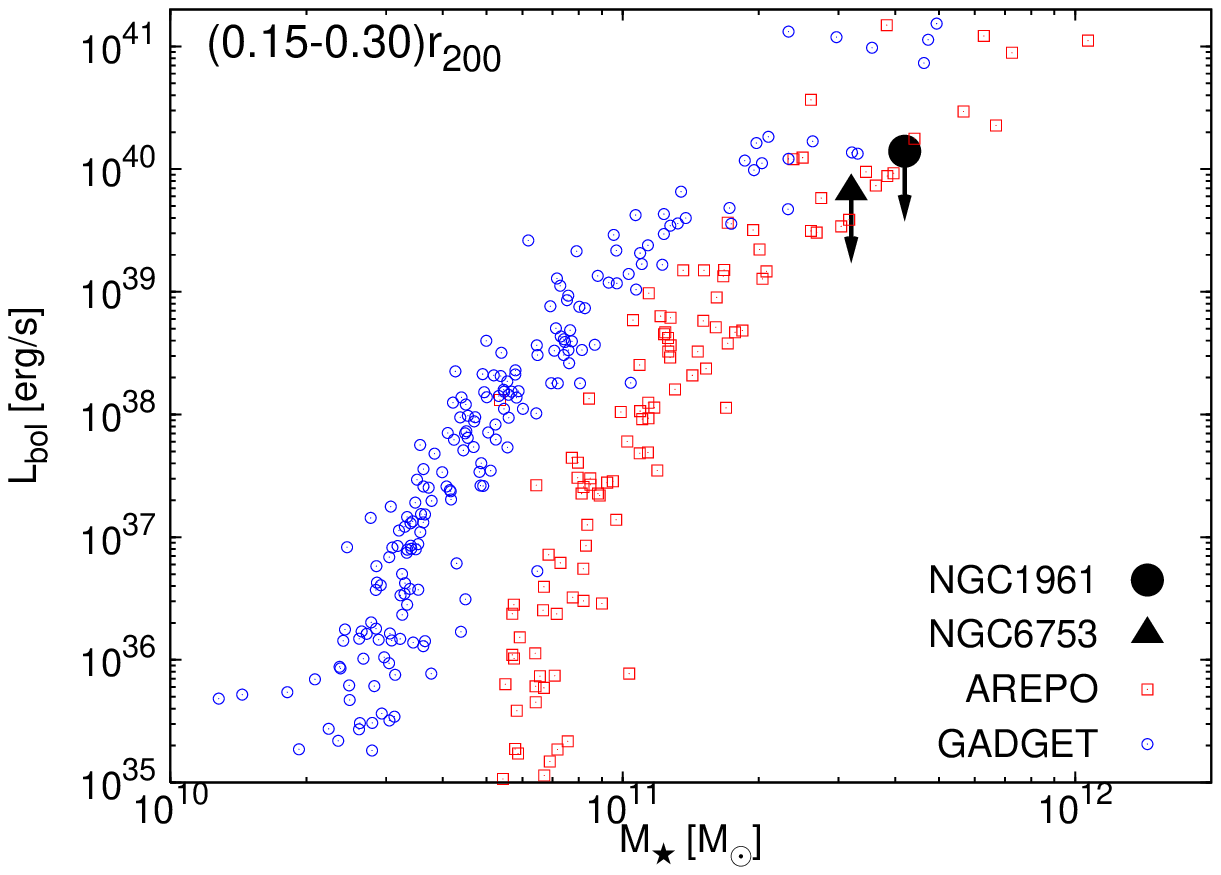}
     \caption{Predicted bolometric X-ray luminosities of the X-ray coronae around disk galaxies as a function of stellar mass for the $(0.05-0.15)r_{200}$ (left panel) and $(0.15-0.30)r_{200}$ radial cuts (right panel). Galaxies simulated by \textsc{arepo} are shown with small boxes, whereas galaxies simulated by  \textsc{gadget} are marked with small circles. The observed X-ray luminosities and $3\sigma$ upper limits for NGC1961 and NGC6753 are shown with the big symbols. The statistical uncertainties on the observed X-ray luminosities are comparable with the size of the symbols.}
\vspace{0.5cm}
     \label{fig:luminosity}
  \end{center}
\end{figure*}

\textsc{gadget} is a widely used and well tested SPH code, which  simultaneously conserves energy and entropy despite the use of fully adaptive smoothing lengths \citep[][]{springel02}. The gravity calculation is split into short-range and long-range components, where the short-range forces are calculated with a hierarchical oct-tree algorithm
\citep{barnes86,hernquist87} and the long-range forces are evaluated with a particle mesh (PM) method \citep[e.g.][]{hockney81}. The simulations presented here are based on a default setting of the numerical SPH parameters \citep[see][for details]{vogelsberger12}. \textsc{arepo} is a second-order accurate finite volume method that solves the Euler equations using piece-wise linear reconstruction and a calculation of hydrodynamical fluxes at each cell interface with an exact Riemann solver. The basic solution strategy of the code is that of the well-known MUSCL-Hancock scheme. \textsc{arepo} employs an unstructured mesh based on a Voronoi tessellation using a set of mesh-generating points. This mesh is allowed to move freely as a function of time. In the simulations presented here, the motion of the mesh-generating points is tied to the hydrodynamical flow resulting in a quasi-Lagrangian character and makes the mesh automatically adaptive. Such a scheme combines the advantages of classical adaptive mesh refinement (AMR) and SPH codes. For example, shocks, contact discontinuities, and mixing are  significantly better resolved in \textsc{arepo} than in classical SPH codes like \textsc{gadget}. 

The simulations discussed here adopt the cosmological parameters $\Omega_{\rm{m,0}}=0.27$, $\Omega_{\rm{{\Lambda,0}}}=0.73$, $\Omega_{\rm{{b,0}}}=0.045$, $\sigma_{\rm{8}}=0.8$, $n_{\rm{s}}=0.95$ and $H_{\rm{0}}=100\,h\,{\rm km}\,{\rm s}^{\rm -1}\,{\rm Mpc}^{\rm -1}= 70\,{\rm km}\,{\rm s}^{\rm -1}\,{\rm Mpc}^{\rm -1}$. These parameters are consistent with the most recent WMAP-7 measurements \citep[][]{komatsu11}.  The simulation covers a volume of $(20\,h^{-1}\,{\rm Mpc})^3$ sampled by $2 \times 512^3$ dark matter particles/cells with a mass of $3.722 \times 10^6 \ \rm{M_{\odot} \ h^{-1}}$, the gravitational softening length is $1\,h^{-1}\,{\rm kpc}$ comoving.  The simulations contain also $512^3$ SPH particles, and the AREPO (de-)refinement is setup such that the total number of Voronoi cells stays roughly constant also at $512^3$. This results in a SPH particle/cell mass of $7.444 \times 10^5 \ \rm{M_{\odot} \ h^{-1}}$. Stars inherit their mass from SPH particles/cells, i.e. the mass of stellar particles is of the same order. For further details on the simulation setup we refer to \citet{vogelsberger12}.

The \textsc{gadget} and \textsc{arepo} simulations follow the same physics. In particular, the simulations consist of a collisionless dark matter fluid and an ideal baryonic gas augmented with additional terms that account for radiative processes (optically-thin radiative cooling of a primordial mixture of helium and hydrogen plus a uniform, time-dependent ionising UV background) and star formation.  Gas can collapse to high density and turn into stars.  Both codes describe this process with the star formation and supernova feedback model introduced in \cite{springel03} assuming a Salpeter IMF.

\begin{figure*}[!t]
  \begin{center}
    \leavevmode
      \epsfxsize=8.7cm\epsfbox{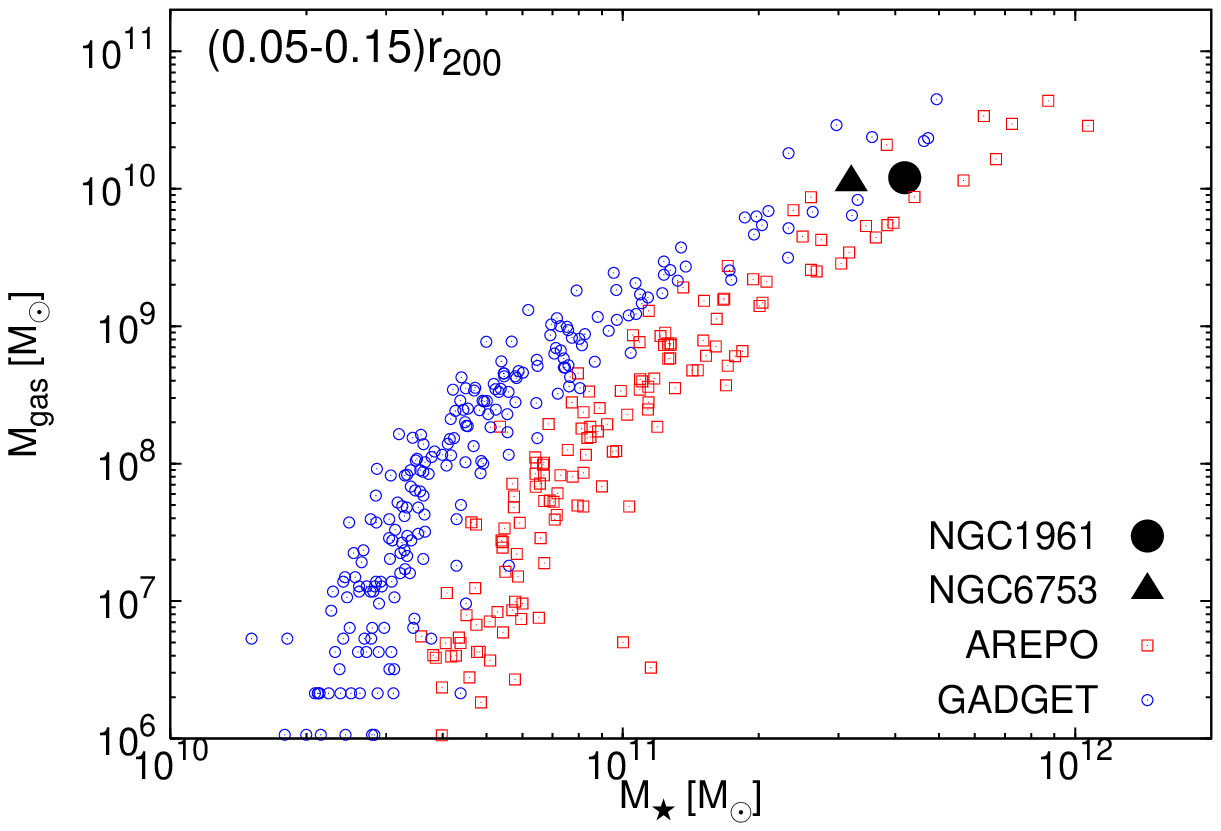}
\hspace{0.1cm}
      \epsfxsize=8.27cm\epsfbox{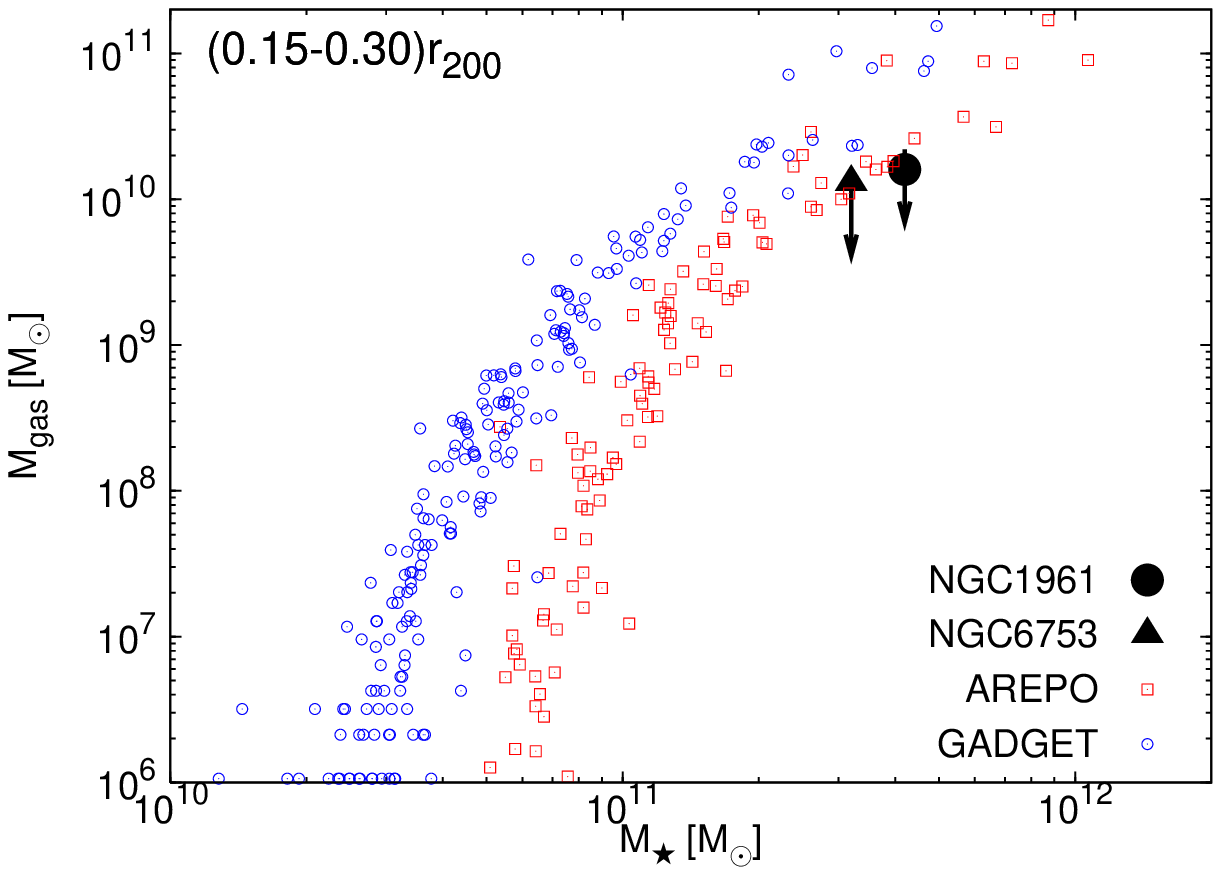}
     \caption{Predicted hot gas mass of the  X-ray coronae around disk galaxies as a function of stellar mass for the $(0.05-0.15)r_{200}$ (left panel) and $(0.15-0.30)r_{200}$ regions (right panel). Only gas particles with temperature above $0.1$ keV were used to compute the gas mass in the simulations. Galaxies simulated by the \textsc{arepo} code are shown with small boxes, whereas galaxies simulated by the \textsc{gadget} code are marked with small circles. The measured hot gas masses and $3\sigma$ upper limits for NGC1961 and NGC6753 are shown with the big symbols. The statistical uncertainties on the hot gas masses are comparable with the size of the symbols.}
\vspace{0.5cm}
     \label{fig:gasmass}
  \end{center}
\end{figure*}

A comparison between the moving mesh code (\textsc{arepo}) and the SPH code (\textsc{gadget}) has been performed recently in a series of papers \citep{vogelsberger12,keres12,sijacki12,torrey11,bird12}.  In their study, \citet{vogelsberger12} recognized a number of important differences between the two codes. Most importantly, significantly more gas cools out of halos at low redshifts in \textsc{arepo} than in \textsc{gadget}, thereby yielding more massive and more extended stellar disks. This difference can be attributed to the stronger cooling flows that evolve in \textsc{arepo} at low redshifts, yielding larger star-formation rates. At the same time, however, the hot X-ray coronae around massive galaxies are less dense, hence less luminous and have lower mass in \textsc{arepo} than in  \textsc{gadget}. Additionally, the luminosity weighted gas temperatures are somewhat lower in \textsc{arepo} than in \textsc{gadget}. The hotter gas temperatures and the weaker gas cooling in the \textsc{gadget} code can be explained by the different dissipative heating rates. Indeed, \textsc{arepo} produces higher dissipative heating in the halo infall regions, and \textsc{gadget} produces higher dissipation in the outer regions of halos. \citet{vogelsberger12} argue that the higher dissipation in \textsc{gadget} is likely to be of spurious nature \citep[see also][]{bauer12}. Thus, the differences described above between the \textsc{arepo} and \textsc{gadget} simulations are purely due to the  inaccuracies of the  hydrodynamic solver. 

Given the above differences between the \textsc{arepo} and \textsc{gadget} simulations, it is essential to observationally test these models. Confronting the observed and predicted properties of X-ray coronae around massive spiral galaxies is advantageous for the following two reasons. First, since the existence of hot gaseous coronae around massive disk galaxies is a basic prediction of galaxy formation models \citep{white78}, disk galaxies, such as NGC1961 and NGC6753, offer a clean observational test of these models. Second, the extended X-ray coronae are well-resolved in numerical simulations, hence their evolution can be followed through a large redshift range. Thus, in the following sections we compare the observed properties of hot gasesous coronae around NGC1961 and NGC6753 with those predicted by the \textsc{arepo} and the \textsc{gadget} simulations.

\subsection{X-ray luminosities}
\label{sec:luminosity}
As a first probe, we confront the observed X-ray luminosities around NGC1961 and NGC6753 with those predicted by the \textsc{arepo} and \textsc{gadget} simulations. To facilitate the comparison with present-day and future simulations, all radial distances are converted to $r_{200}$ units, where $r_{200}$ is defined as the virial radius. The $r_{200}$ radii of NGC1961 and NGC6753 are $\sim$$470$ kpc and $\sim$$440$ kpc, respectively (Section \ref{sec:galaxies}).  Given the relatively small volume of the simulations, we did not make a specific selection of spiral galaxies. However, the studied volume mainly consists of spirals since the simulations do not include the physics required to form a significant population of ellipticals (for details see Section \ref{sec:feedback}). We note that the simulated galaxies are not members of rich galaxy clusters.

For both NGC1961 and NGC6753, we compare the observed and predicted bolometric X-ray luminosities in two circular annuli with radii of $(0.05-0.15)r_{200}$ (inner region) and $(0.15-0.30)r_{200}$ (outer region). Note that the central regions ($<0.05r_{200}$)   are excluded due to the various contaminating factors associated with the stellar body of the galaxies (Section \ref{sec:introduction}). In the $(0.05-0.15)r_{200}$ region, the X-ray luminosities are obtained  from the best-fit spectra. In the $(0.15-0.30)r_{200}$ region,  in the absence of X-ray gas detection, we place $3\sigma$ upper limits on the luminosities using adopted gas parameters (Section \ref{sec:spectra}). The bolometric X-ray luminosities and upper limits for NGC1961 and NGC6753 are listed in Table \ref{tab:luminosity}.

\begin{figure*}[!t]
  \begin{center}
    \leavevmode
      \epsfxsize=8.7cm\epsfbox{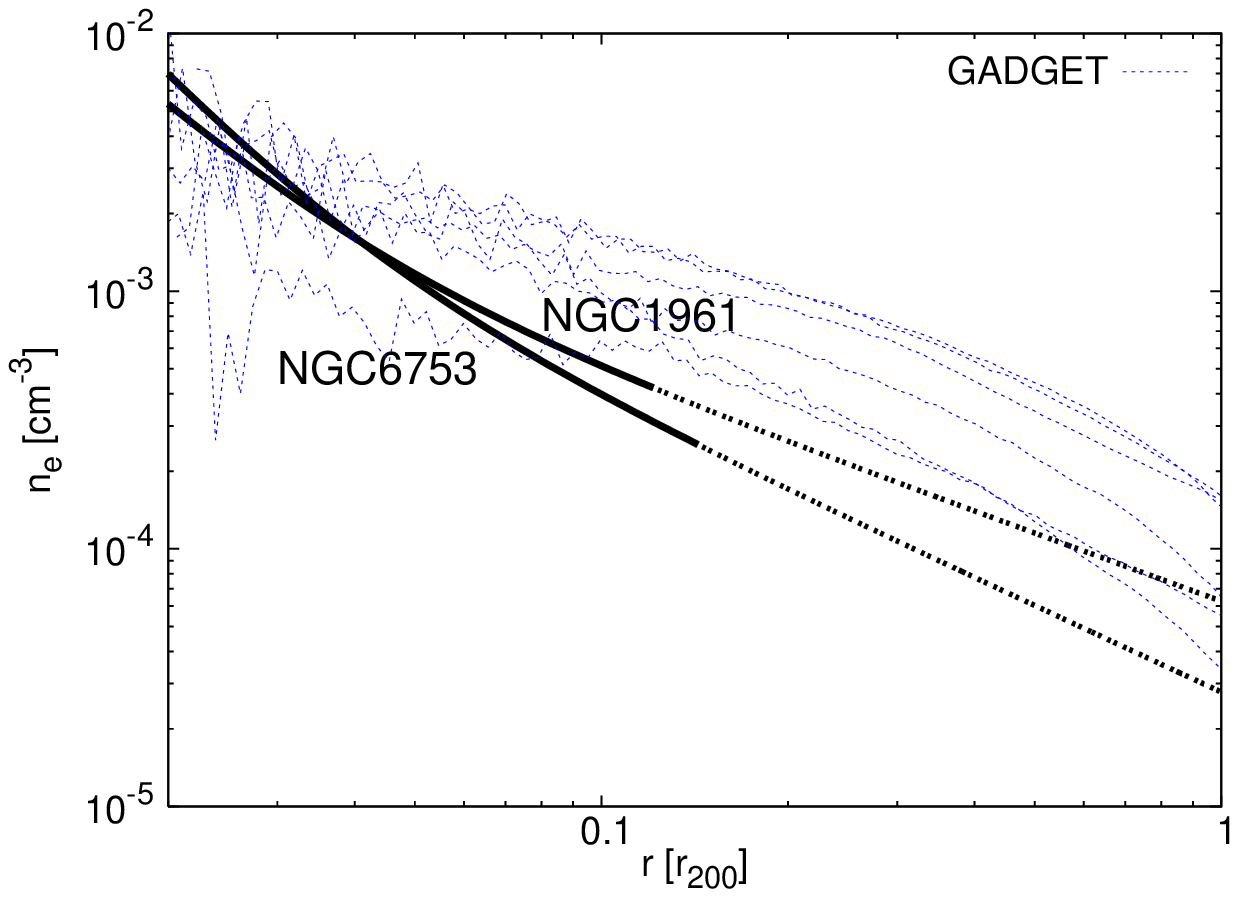}
\hspace{0.1cm}
      \epsfxsize=8.7cm\epsfbox{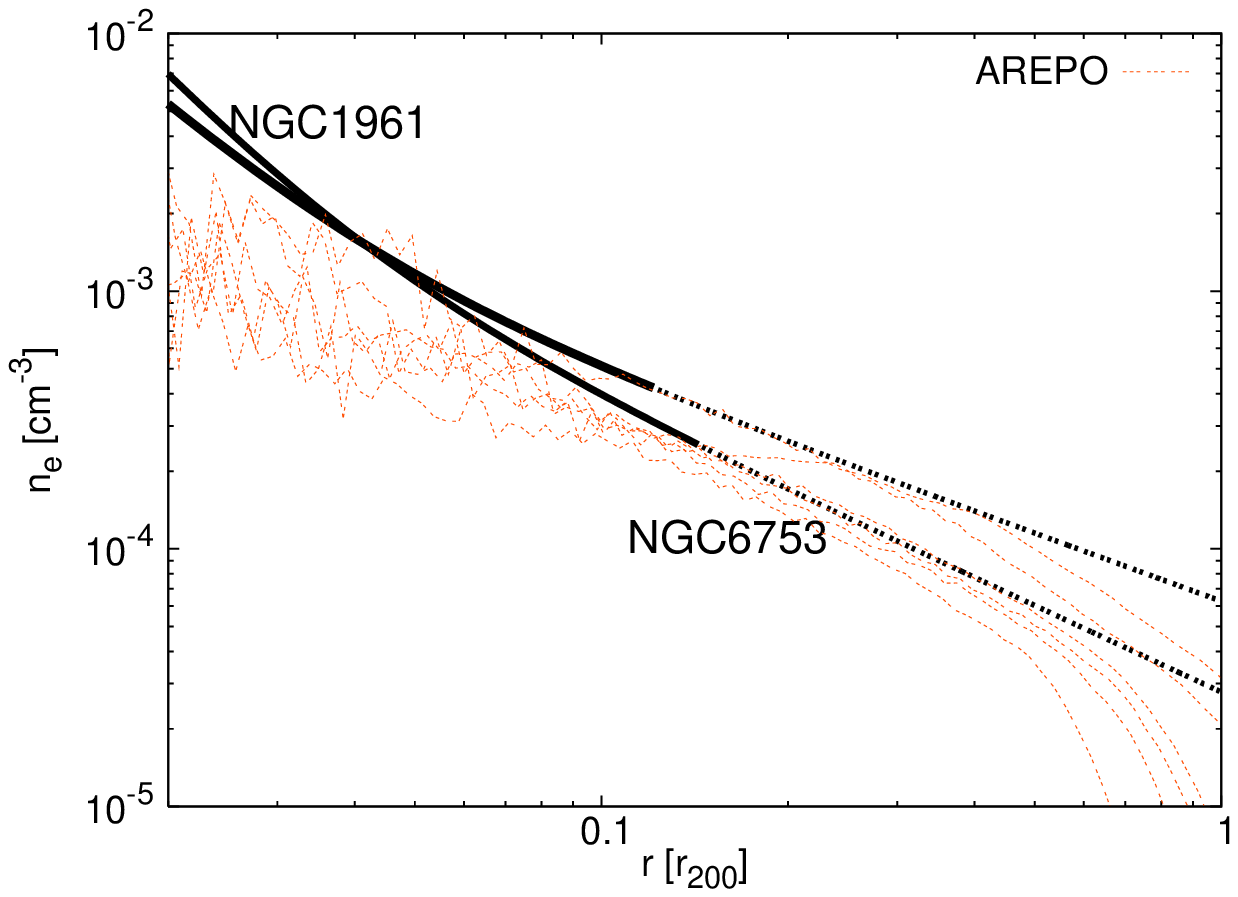}
     \caption{Observed and predicted electron density profiles as the function of the radius for the \textsc{gadget} (left panel) and \textsc{arepo}  (right panel) simulations. Note that the radius is measured in units of $r_{200}$.  The thick solid lines show the observed density profiles, whereas the short dashed thick lines show the extrapolated density profiles based on the best-fit $\beta$-models  (Section \ref{sec:gas_mass}). The statistical uncertainties on the observed profiles are comparable with the difference between the two profiles. The predicted density profiles originate from six representative simulated galaxies, whose stellar masses are comparable with those of NGC1961 and NGC6753. }
\vspace{0.5cm}
     \label{fig:density}
  \end{center}
\end{figure*}

In Figure \ref{fig:luminosity} we show the predicted bolometric X-ray luminosites as a function of the stellar mass for the $(0.05-0.15)r_{200}$ and $(0.15-0.30)r_{200}$ regions. In these plots we mark the corresponding observed values and upper limits for NGC1961 and NGC6753. It is apparent from Figure \ref{fig:luminosity} that, on average, the predicted X-ray coronae are more luminous in \textsc{gadget} than in \textsc{arepo}. Although this difference is clear in the inner region, it becomes even more evident in the outer regions. As a caveat, we note that above $\sim$$2\times10^{11} \ \rm{M_{\odot}}$   this difference is less evident, especially in the $(0.05-0.15)r_{200}$ radial range. In the inner region, the observed luminosities for NGC1961 and NGC6753 fall closer to the \textsc{gadget} predictions. However, given the relatively large scatter at the high mass end, no strong conclusions can be drawn. In the outer region, the $3\sigma$ upper limits are approximately consistent with the \textsc{arepo} model predictions, but fall a factor of $\sim$$2-10$ short of the \textsc{gadget} predictions.

\begin{table}
\caption{The best-fit parameters of hot gas density profiles.}
\renewcommand{\arraystretch}{1.3}
\centering
\begin{tabular}{c c c c c c c c}
\hline 
Galaxy&  $n_0$ &$r_c $ & $r_s^{\dagger}$ & $\alpha$ &$\beta$ & $\epsilon^{\dagger}$  & $\gamma^{\dagger}$ \\ 
 & ($ \rm{cm^{-3}}$) & (kpc)  & (kpc) & &   \\ 
\hline
NGC1961 & $1.0\times10^{-3}$ & 18.1 & 400 & 4.40 & 0.29 & 0 &3.0\\
NGC6753 & $1.0\times10^{-3}$ & 16.3 & 400 & 5.37 & 0.37 & 0 &3.0 \\
\hline \\
\end{tabular} 
$^{\dagger}$ These parameters were fixed at the given value.
\label{tab:fit}
\end{table}

\subsection{Hot gas mass and density profiles}
\label{sec:gas_mass}
In Figure \ref{fig:gasmass} we present the predicted gas mass in the $(0.05-0.15)r_{200}$ and $(0.15-0.30)r_{200}$ regions as a function of the stellar mass. In both \textsc{arepo} and \textsc{gadget} simulations we only consider the hot gas components, whose temperature is above $0.1$ keV.  The observed gas masses and upper limits are given  in Table \ref{tab:values}. 

Figure \ref{fig:gasmass} demonstrates that \textsc{gadget} predicts more massive X-ray coronae than \textsc{arepo}. This difference is significantly larger in the outer region, thereby suggesing that the density profiles of the hot gas coronae fall more rapidly in the \textsc{arepo} than in \textsc{gadget} simulations. In the inner region, the observed gas mass is somewhat higher than that predicted by the moving mesh code but falls short of the SPH simulations. In the outer region, the $3\sigma$ upper limits on the gas mass are broadly consistent with the \textsc{arepo} predictions. However, the \textsc{gadget} simulation overpredicts the gas mass with at least factor of 2, but for most simulated galaxies with comparable stellar mass the predicted gas mass exceeds the $3\sigma$ upper limits with factors of about $8-10$.

To further probe the structure formation models, we compare the observed and predicted electron density profiles, which allow a more detailed comparison between the observed and simulated gas distributions. Hence not only the volume averaged quantities can be confronted but also the normalization and the shape of the density profiles will be tested. Since the 
hot gas is only detected to $\sim$$0.15r_{200}$, we extrapolate our measurements out to the virial radius.  To this end, we describe the observed surface brightness distribution in the $(0.02-0.15)r_{200}$ region with a modified $\beta$-model. Under the assumption of  constant metallicity gas, we derive the emission measure profile at a projected radius ($r$) following \citet{vikhlinin06}:
\begin{eqnarray}
 n_{p} n_{e}=n_0^2 \frac{(r/r_c)^{-\alpha}}{(1+r^2/r_c^2)^{3\beta-\alpha/2}} \frac{1}{(1+r^\gamma/r_s^{\gamma})^{\epsilon/\gamma}} \ .
\end{eqnarray}
This $\beta$-model consists of a power law-type cusp and also describes the possible steepening of X-ray surface brightness profiles at large  radii \citep{vikhlinin06}. The obtained best-fit parameters of the emission measure profile are listed in Table \ref{tab:fit}. To compare the observed electron density profiles with those produced by  \textsc{arepo} and \textsc{gadget}, we extract the density profiles of six representative galaxies with stellar masses in the range of $(2.5-4.9)\times10^{11} \ \rm{M_{\odot}}$, which is comparable with those of NGC1961 and NGC6753. We stress that the simulated electron density profiles do not correspond to the same halos, hence the simulated \textsc{gadget} and \textsc{arepo} profiles are not directly comparable. This is due to the fact that, while both simulations are affected by the so-called overcooling problem, the stellar mass of galaxies in the \textsc{arepo} simulations is larger than that in the \textsc{gadget} simulations (see Section \ref{sec:feedback} for details). Moreover, given that the virial radii of the simulated galaxies are somewhat different than that of NGC1961 and NGC6753, the density profiles in Figure \ref{fig:density}  cannot be directly compared with Figure \ref{fig:gasmass}, which shows the integrated gas mass as a function of stellar mass.  The selected simulated galaxies have a SFR of $10-20 \ \rm{M_{\odot} \ yr^{-1}}$ at $z=0$ in the \textsc{arepo} simulation (in the \textsc{gadget} it is about a factor of two lower), which is also comparable with the observed values for NGC1961 and NGC6753. As a caveat, we mention that the limited numerical resolution of the simulations implies that halos are not well resolved close to the gravitational softening scale. Therefore, the inner regions ($\lesssim0.05r_{200}$) close to that scale are not sampled by a large number of particles, which causes the more noisy profiles towards the inner parts of the halos. 

\begin{figure}[!t]
  \begin{center}
    \leavevmode
      \epsfxsize=8.7cm\epsfbox{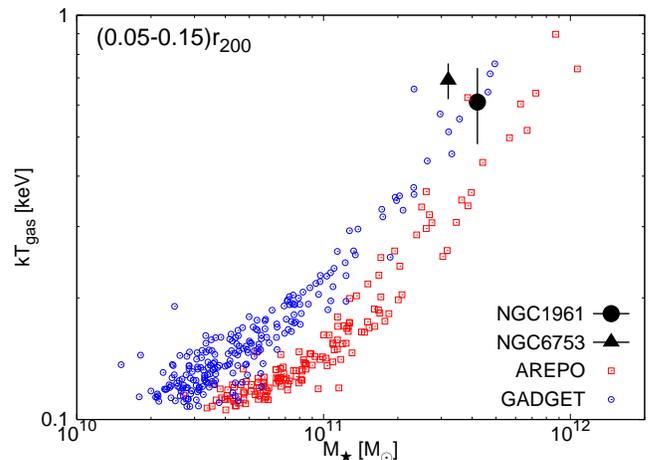}
     \caption{Predicted hot gas temperatures of the X-ray coronae as a function of the stellar mass for the $(0.05-0.15)r_{200}$ range. Simulated galaxies are shown with the small boxes (\textsc{arepo}) and small circles (\textsc{gadget}).  The observed hot gas temperatures for NGC1961 and NGC6753, obtained from the best-fit spectra,   are shown with the big filled symbols.}
\vspace{0.5cm}
     \label{fig:gas_temperature}
  \end{center}
\end{figure}

The comparison between observed and simulated electron density profiles are depicted in Figure \ref{fig:density}. Although the density profiles predicted by \textsc{gadget} show a large scatter, their overall distribution strikingly differs from the observed ones at all radii (left panel of Figure \ref{fig:density}). Indeed, the simulated \textsc{gadget} density profiles  are significantly flatter than those observed (and extrapolated) for NGC1961 and NGC6753, and most of them predict too high densities  in the $(0.05-1)r_{200}$ radial range. Although in the innermost regions  ($<0.05r_{200}$)  the predicted and observed electron densities agree with each other, this result is not conclusive since the innermost regions of galaxies cannot be used to probe galaxy formation models.   At larger radii  ($>0.5r_{200}$) the simulated profiles fall much faster than the extrapolated ones, which, at least partly, may be explained by the absence of efficient supernova and AGN feedback in the model. 

\begin{figure}[!t]
  \begin{center}
    \leavevmode
      \epsfxsize=8.5cm\epsfbox{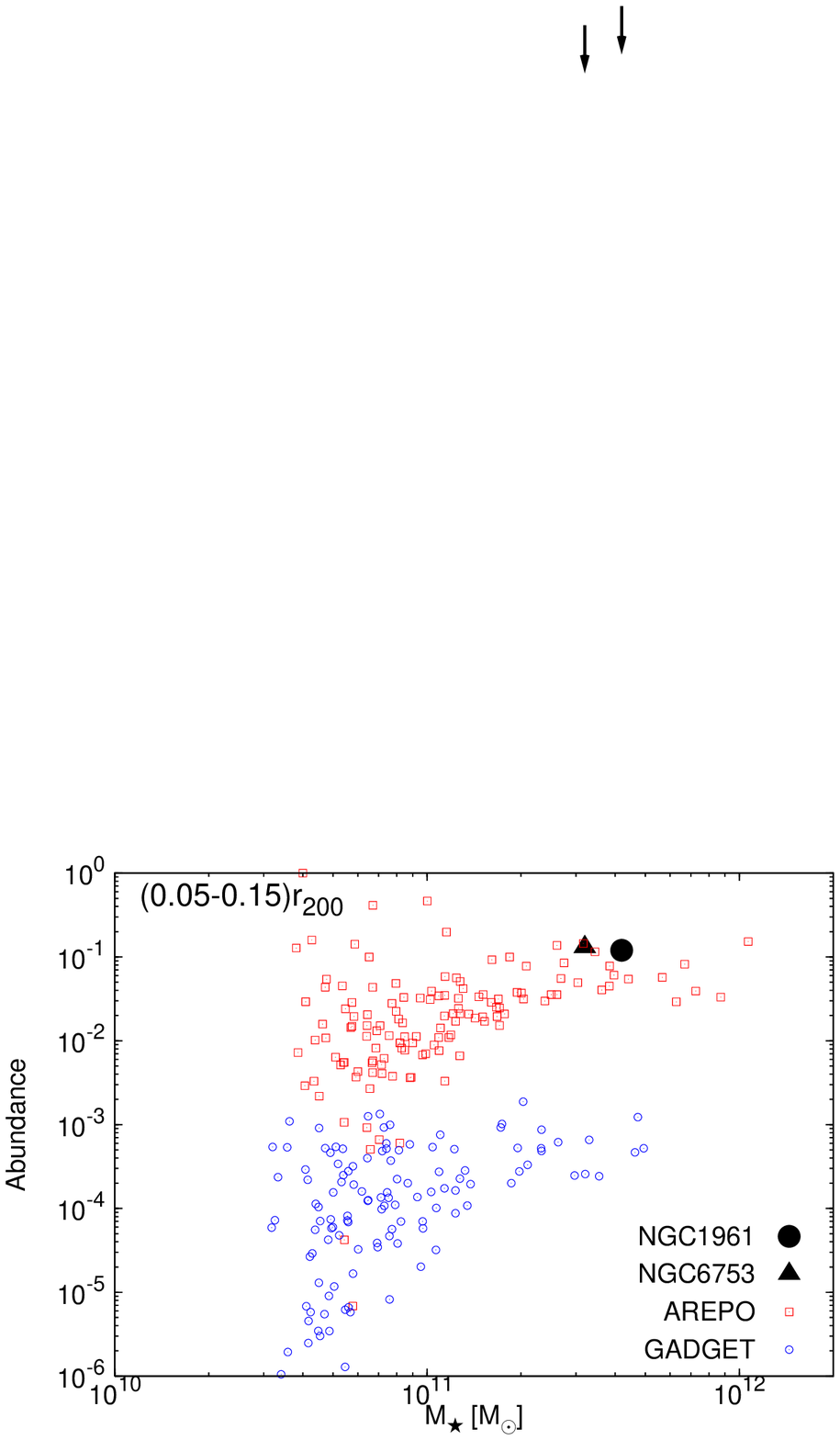}
     \caption{Predicted abundances of the X-ray coronae vs stellar mass for the $(0.05-0.15)r_{200}$ range. Galaxies simulated by  \textsc{arepo} are shown with boxes, and galaxies simulated by \textsc{gadget} are marked with circles. The observed  abundances for NGC1961 and NGC6753 are shown with the big symbols. The statistical uncertainties on the measured values are comparable with the size of the symbols.}
\vspace{0.5cm}
     \label{fig:gas_abundance}
  \end{center}
\end{figure}

The density profiles predicted by the \textsc{arepo} simulations also exhibit a notable scatter (right panel of Figure \ref{fig:density}). Although the predicted densities fall short of the observed values in the innermost regions, the shape of the density profiles and the observed and predicted  $n_e$ values are broadly consistent in the $(0.05-0.5)r_{200}$ range. We mention that within $\sim$$0.05r_{200}$ the X-ray emission is dominated  by gas produced in the course of the stellar evolution of the galaxies, therefore this region does not test the extended X-ray coronae.  Beyond  $\sim$$0.5r_{200}$ the density drops sharper than the extrapolated profiles of NGC1961 and NGC6753. The truncated density profile in the simulation is most likely due to the absence of efficient supernova and AGN feedback in the models. Implementing efficient energy feedback will likely  result in more extended gas distributions, resulting in a better agreement with the observed profiles at large galactocentric radii.

Based on the density profiles we conclude that (1) \textsc{arepo} describes the observed density profiles better than \textsc{gadget} at radii $<0.5r_{200}$; and (2) at larger radii both simulations fail to reproduce the shape of the extrapolated profiles, presumably due to the absence of efficient feedback in the models.

\subsection{Hot gas temperature}
\label{sec:temperature}
In Figure \ref{fig:gas_temperature} we depict the predicted hot gas temperatures in the $(0.05-0.15)r_{200}$ radial range as the function of the stellar mass. The predicted temperatures are luminosity-weighted values.

In general,  \textsc{gadget} predicts higher gas temperatures than \textsc{arepo} throughout the entire stellar mass range. For both simulations, the expected temperatures are in the relatively narrow  $kT=0.3-0.7$ keV range for galaxies with $\sim$$4\times10^{11} \ \rm{M_{\odot}}$. For NGC1961 and NGC6753 the observed gas temperatures are $kT\sim0.6-0.7$ keV, which places them at the high end of the predictions. The observed values are in fair agreement with the SPH simulation and exceed the values predicted by the moving mesh code. 

However, at present, neither simulation code implements efficient energy feedback from supernovae and supermassive black holes, which  serves as an additional heating source. Therefore,  including the efficient feedback in the simulations will yield  higher gas temperatures, and possibly better agreement with the observed values. We note that based on the presently available X-ray data, the temperature distribution of the hot gas cannot be studied in detail. To better explore the temperature structure of X-ray coronae, further deep observations are required.

\subsection{Abundances of the hot gas}
As a final test, we investigate the observed and predicted abundances of the hot X-ray coronae in the $(0.05-0.15)r_{200}$ radial cut. The observed abundances for NGC1961 and NGC6753 are  the best-fit values from the spectral fitting procedure, which  are given in Table \ref{tab:values}. The predicted values are luminosity-weighted abundances.

In Figure \ref{fig:gas_abundance} we show the predicted abundances for the \textsc{arepo} and \textsc{gadget} simulations as a function of  stellar mass. The observed    values for NGC1961 and NGC6753 are overplotted. For the two structure formation simulations, there is a striking difference between the predicted abundances. Whereas for galaxies with a few $\times10^{11} \ \rm{M_{\odot}}$ stellar mass, \textsc{arepo} predicts $\sim$$0.1$ Solar abundance.  The expected values in \textsc{gadget} are about two orders of magnitude less, $\sim$$10^{-3}$ Solar. The observed values for NGC1961 and NGC6753 are $\sim$$0.1$ Solar, which are in good agreement with that predicted by \textsc{arepo}. We thus conclude that the observed abundances of the hot gas around NGC1961 and NGC6753 strongly favor the moving mesh code over the SPH simulations. 

The abundance difference between the codes is caused by two effects. First, due to the increased cooling, \textsc{arepo} exhibits higher star formation rates than \textsc{gadget}, which is attributed to the different dissipative heating rates (Section \ref{sec:codes}). Since metal production in the simulations is tied to star formation assuming a fixed yield, the higher star formation in \textsc{arepo} automatically leads to a production of higher abundances. Second, the mixing is more  efficient in  \textsc{arepo} than in \textsc{gadget}. This point is related to the intrinsic mixing problems of standard SPH codes.  In the  \textsc{gadget} simulations, metals are transported by individual SPH particles and they do not undergo any explicit metal diffusion process. Contrarily, the metals are advected and can mix well within the gas in the \textsc{arepo} code. Accordingly, metals stay closer to the central part of galaxies and haloes in \textsc{gadget} compared to \textsc{arepo}. This difference becomes particularly apparent for more massive galaxies, where enrichment of halos through mixing is highly ineffective in \textsc{gadget} (Figure \ref{fig:gas_abundance}). In massive galaxies the abundances differ by about two orders of magnitude between the two simulations, but this difference gets progressively lower in smaller mass systems. The mixing of metals in the \textsc{arepo} code will be further discussed in the upcoming work by Vogelsberger et al. (2013, in preparation). 
Finally, we note that the inclusion of explicit supernova winds is another source of halo gas enrichment, which is not included in our simulations.

\begin{figure}[!b]
  \begin{center}
    \leavevmode
      \epsfxsize=8.7cm\epsfbox{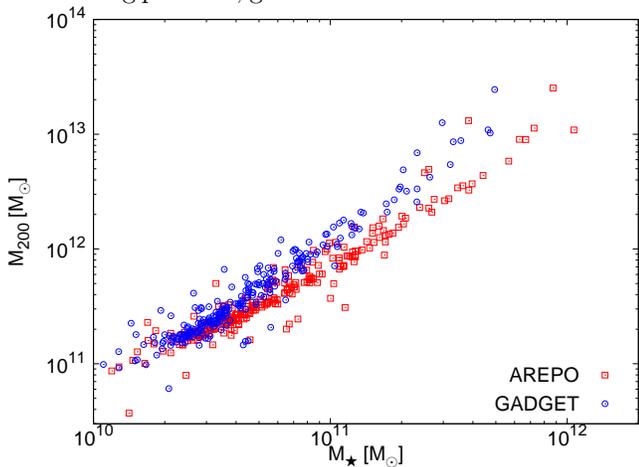}
      \caption{Virial mass as a function of stellar mass for the \textsc{arepo} and \textsc{gadget}  simulations. Galaxies simulated by the \textsc{arepo} code are shown with boxes, whereas galaxies simulated by the \textsc{gadget} code are marked with circles. Note that both simulations suffer by the so-called overcooling problem, that is too much gas cools out of the halos, thereby producing too massive stellar disks relative to their dark matter halos. The stellar masses of NGC1961 and NGC6753 are $4.2\times10^{11} \ \rm{M_{\odot}}$ and $3.2\times10^{11} \ \rm{M_{\odot}}$, respectively.}
\vspace{0.5cm}
     \label{fig:overcooling}
  \end{center}
\end{figure}

\subsection{Limitations of the present simulations}
\label{sec:feedback}
Although AGN feedback and galactic winds are believed to have a profound impact on the evolution of galaxies, the  galaxy formation simulations discussed in this paper do not incorporate efficient energy feedback from AGN and supernovae.  The impact of AGN feedback is observed in different scales in the local Universe: in galaxies \citep[e.g.][]{finoguenov01,jones02,kraft11}, galaxy groups \citep[e.g.][]{david09,gitti10,randall11}, and galaxy clusters \citep[e.g.][]{fabian00,forman05,nulsen05,bogdan11a}. Moreover, theoretical studies also point out that the AGN  feedback is capable of heating the cool gas and expelling it to larger radii, thereby quenching the star-formation and regulating the growth of supermassive black holes \citep{silk98,dimatteo05,sijacki07}. Galactic winds driven by the energy input of supernovae eject metal-rich gas to large radii, thereby enriching the extraplanar gas \citep{oppenheimer06}. Additionally, galactic winds play a major role in reproducing the cosmic star-formation history at high redshifts \citep{schaye10}. Furthermore, implementing efficient AGN and supernova feedback  will address one of the major shortcomings of the present models: the so-called overcooling of galaxies. Without efficient feedback  too much gas cools out of the halos both in the \textsc{arepo} and the \textsc{gadget} simulations, hence produce overly massive stellar components relative to their dark matter halos. Although both simulation codes are affected by the overcooling problem, galaxies in \textsc{arepo}  are more massive  than their counterparts in  \textsc{gadget} (Figure \ref{fig:overcooling}). 

The present models also do not incorporate the contribution of metal cooling to the emission.  However, the cooling of metal species plays a significant role at the observed sub-keV temperatures \citep{voort11}. Indeed, metal cooling notably influences the   temperature structure, hence the X-ray luminosity of the hot gas. Metal line cooling also increases the star formation rates, and affects the build-up of galaxies. On the other hand, AGN and supernova feedback typically heat the gas and decrease the star formation rates, such that missing feedback and missing metal line cooling partially cancel each other.

In subsequent papers we will investigate structure formation simulations, which incorporate the above discussed missing physics. Thus, the forthcoming studies will lead to a physically more realistic comparisons between the observed X-ray coronae and numerical models.

\subsection{Synopsis about the \\ \textsc{arepo} and \textsc{gadget} simulations}
In this section we compared the observed physical properties of hot X-ray coronae around two  massive spiral galaxies with those predicted by two structure formation simulations. The moving mesh code (\textsc{arepo}) and the SPH based code (\textsc{gadget}) uses fully identical subresolution physics and gravity solver, but they employ different methods to solve the equations of hydrodynamics. Based on the comparison we reached the following major conclusions:
\begin{enumerate}
\item In the $(0.05-0.15)r_{200}$ radial range the X-ray luminosities and hot gas masses broadly agree with both simulations. However, in the $(0.15-0.30)r_{200}$ region the upper limits for  X-ray luminosities and gas masses are more consistent with \textsc{arepo}, since \textsc{gadget} overpredicts both quantities.  
\item The observed (and extrapolated) density profiles in the $(0.05-0.5)r_{200}$ radial range are in fair agreement with those predicted by \textsc{arepo}. Contrarily, \textsc{gadget} predicts markedly different density profiles, which disagree with the shape and normalization of the observed and extrapolated profiles in the $(0.05-1)r_{200}$ region. 
\item The observed abundances in the $(0.05-0.15)r_{200}$  region are in  good agreement with those predicted  by \textsc{arepo}, but exceed the predictions of \textsc{gadget} by about two orders of magnitude.
\end{enumerate}

Although the above listed points suggest that the moving mesh code gives a significantly better description of structure formation, further  observational and theoretical efforts are required before definite conclusions can be drawn. On the observational side, it is necessary to further push the limits of X-ray observations, and probe the X-ray coronae at even larger radii. Such observations would allow the more precise comparison of observations and numerical models. Moreover, the parameter space needs to be better explored, that is the X-ray coronae of other, preferably lower mass, galaxies should be observed and characterized. On the theoretical side, efficient energy feedback from AGN and supernovae must be included along with cooling from metal species. The comparison with physically more realistic models will lead to a significantly better understanding of the physical processes, which influence galaxy formation models.

\begin{table*}[!t]
\caption{Mass budget of NGC1961 and NGC6753.}
\begin{minipage}{18cm}
\renewcommand{\arraystretch}{1.3}
\centering
\begin{tabular}{c c c c c c c c}
\hline 
Galaxy & $M_{\star}$ & $M_{\rm{gas,hot}}$  &  $M_{\rm{gas,cold}}$  & $M_{\rm{group}}$  & $M_{\rm{b,tot}}$ & $M_{\rm{DM}}$ & $f_{\rm{b}}$ \\ 
& ($M_{\odot}$) & ($M_{\odot}$) & ($M_{\odot}$) &  ($M_{\odot}$) & ($M_{\odot}$) & ($M_{\odot}$) & \\
& (1) &  (2)  &   (3)  &  (4) &  (5)  &  (6) &  (7)   \\
\hline
NGC1961  &$4.2\times10^{11}$  &  $9.4\times10^{11}$ & $4.7\times10^{10}$ & $1.4\times10^{11}$ &  $1.55\times10^{12}$ & $1.2\times10^{13}$ & $0.11$ \\
NGC6753  & $3.2\times10^{11}$ &  $3.9\times10^{11}$ &  $2.0\times10^{10}$ & $2.7\times10^{11}$ &  $1.00\times10^{12}$& $1.0\times10^{13}$ & $0.09$ \\

\hline \\
\end{tabular} 
\end{minipage}
\textit{Note.} Columns are as follows. (1) Total stellar mass of the sample galaxies. (2) Estimated hot X-ray gas mass within the virial radius, obtained from the extrapolation of the $\beta-$models (Section \ref{sec:massbudget}). (3) Cold gas mass. (4) Total stellar mass of group members. (5) Total baryon mass, which is the sum of columns $(1)-(4)$. (6) Estimated mass of the dark matter halo. (7) Baryon mass fraction, defined as $f_{\rm{b}}=M_{\rm{b,tot}}/(M_{\rm{DM}}+M_{\rm{b,tot}})$. \\
\label{tab:masses}
\end{table*}

\section{Discussion}
\label{sec:discussion}
\subsection{The role of starburst driven winds}
Although the goal of the present work is to explore the hot gaseous X-ray coronae   around massive spiral galaxies, it is feasible that a certain fraction of the detected hot gas was uplifted from the galactic disks. In principle, galaxies with sufficiently high specific star-formation rates are capable of driving starburst driven winds. Based on theoretical computations, \citet{strickland04b} estimated that supernova blowout occurs at the (surface area) specific SN rate $\gtrsim$$25 \ \rm{SN \ Myr^{-1} \ kpc^{-2}}$, where the  surface area  is  derived as the square of the $D_{25}$ diameter. In fair agreement with the theoretical computations, \citet{strickland04b} detect extraplanar X-ray emission in galaxies, whose specific SN rates are $\gtrsim$$40 \ \rm{SN \ Myr^{-1} \ kpc^{-2}}$.

To probe the importance of SN driven winds in NGC1961 and NGC6753, we compute their area specific SN rates. Following \citet{heckman90}, the SN rate is derived from $ R_{\rm{SN}} = 0.2 L_{\rm{FIR}}/10^{11} \ \rm{L_{\odot}}$, where $L_{\rm{FIR}}$ is the total far-infrared luminosity. For NGC1961 and NGC6753 we observe $ L_{\rm{FIR,NGC1961}}=9.0\times10^{10} \ \rm{L_{\odot}} $ and  $ L_{\rm{FIR,NGC6753}}=6.8\times10^{10} \ \rm{L_{\odot}} $  (Section \ref{sec:fir}), which correspond to the SN rates of $R_{\rm{SN,NGC1961}} = 0.18 \ \rm{yr^{-1}}$ and $R_{\rm{SN,NGC6753}} = 0.14 \ \rm{yr^{-1}}$, respectively. Based on these values and the  $D_{25}$ major axis diameters, we estimate  area specific SN rates of $\sim$$36 \ \rm{Myr^{-1} \ kpc^{-2}}$ for NGC1961 and $\sim$$97 \ \rm{Myr^{-1} \ kpc^{-2}}$ for NGC6753. The value obtained for NGC1961 is comparable albeit somewhat lower than the observational lower limit required for SN blowout. In NGC6753 the specific SN rate is higher than the critical rate for SN blowout, suggesting that a fraction of the detected warm gas may be uplfited from the disk. 

Despite the relatively high specific SN rates, the following four arguments suggest that the  detected X-ray coronae around NGC1961 and NGC6753 cannot be entirely attributed to starburst driven winds.  1) The morphology of the X-ray gas (Figures \ref{fig:img_1961}, \ref{fig:img_6753}, \ref{fig:wedges}) is fairly symmetric. This indicates that the  hot  gas resides in  hydrostatic equilibrium in the potential well of the galaxies, and is not in a bipolar starburst driven outflow. 2) Although the specific SN rates are comparable or exceed the criteria for SN blowout, the derived values are signifcantly lower than that observed in starburst galaxies, for example in M82 ($\sim$$1200 \ \rm{M_{\odot} \ Myr^{-1} \ kpc^{-2}}$) or NGC253 ($\sim$$180 \ \rm{M_{\odot} \ Myr^{-1} \ kpc^{-2}}$) \citep{strickland04b}. 3) The extraplanar X-ray emission in starburst galaxies is confined to within $\sim$$15$ kpc \citep{strickland04a}, whereas we detect hot gas out to $\sim$$60$ kpc around NGC1961 and NGC6753. Additionally, we emphasize that the central $0.05r_{200}$ ($\sim$$23$ kpc) regions are excluded when studying the properties of the diffuse X-ray emission. 4) The cooling time of the hot gas is on the time scale of $\sim$$50$ Gyrs in the inner region. Therefore, the hot gas can be treated as quasi-static gas as opposed to a starburst driven wind.  Taken together, this evidence suggests that only fraction of the warm gas around NGC1961 and NGC6753 could originate from starburst driven winds. 

The above considerations demonstrate that \textit{at the present epoch} the X-ray appearance of NGC1961 and NGC6753 is not determined by starburst driven winds. However, at higher redshifts gas-rich galaxy mergers occured more frequently, which  could have triggered  more active phases, possibly leading to  starburst driven winds. Such winds do not only expel a notable amount of gas to larger radii, but also play a major role in enriching the hot extraplanar gas, which is reflected by the non-zero abundance of the hot gas around NGC1961 and NGC6753. Thus, at some point of their evolution, the hot X-ray coronae are likely to be influenced by starburst driven winds.

\subsection{Baryon mass fraction}
\label{sec:massbudget}
The detection of hot X-ray coronae around NGC1961 and NGC6753 allows us to estimate their baryon budget  and derive their baryon mass fraction. To account for the baryon content, we consider the total hot X-ray gas mass, the stellar mass of the galaxies, the cold gas mass, and the stellar mass of the group member galaxies. The detailed mass budget is given in  Table \ref{tab:masses}. 

In massive galaxies, such as NGC1961 and NGC6753, the hot X-ray emitting gas adds a notable contribution to the baryon budget. Therefore, we derive the confined hot gas mass by extrapolating the density profiles out to the virial radius (Section \ref{sec:gas_mass}).  For NGC1961 and NGC6753 the obtained total hot gas masses within  $r_{200}$ are $9.4\times10^{11} \ \rm{M_{\odot}}$ for NGC1961 and $3.9\times10^{11} \ \rm{M_{\odot}}$, respectively. We emphasize that these numbers should only be considered as a crude estimate due to the following systematic and statistical uncertainties: 1) The major source of uncertainty originates from the unexplored nature of the hot gas  beyond $0.15r_{200}$ radius. Indeed, a temperature and/or abundance gradient could drastically alter the extrapolated density profiles, and hence the total gas mass within the virial radius. 2)  The hot gas may not have a spherically symmetric distribution at large radii. 3) The applied $r_{200}$ radii may deviate from the real $r_{200}$ radii. 4) Statistical uncertainties in the best-fit parameters of the $\beta$-model. 5) Systematic uncertainties in the derived mass-to-light ratios.

The stellar mass is derived from the 2MASS K-band images and the K-band mass-to-light ratios (Section \ref{sec:2mass}).  For the cold gas mass we use the HI mass of $4.7\times10^{10} \ \rm{M_{\odot}}$ for NGC1961 \citep{haan08}, and we estimate the HI mass of  about $2\times10^{10} \ \rm{M_{\odot}}$ for NGC6753 \citep{haynes84}. Since both galaxies are located in poor groups \citep{fouque92}, the baryon mass of these galaxies should also be included in the baryon budget. The baryon mass of the group members is dominated by their stellar mass, which we estimate based on their K-band luminosity, using $M_{\star}/L_{K}=0.8$, and assuming that they lie at the same distance as NGC1961 and NGC6753. Finally, we mention that within the \textit{XMM-Newton} FOV we do not detect hot intragroup medium around the sample galaxies. For NGC1961 this conclusion is confirmed by the archival \textit{Chandra} observations \citep{anderson11}. 

The  total baryonic masses of NGC1961 and NGC6753 are $M_{\rm{b,tot}} = 1.55\times10^{12} \ \rm{M_{\odot}}$ and $ M_{\rm{b,tot}} = 1.00\times10^{12} \ \rm{M_{\odot}}$, respectively (Table \ref{tab:masses}). The estimated dark matter halo mass of the galaxies are $M_{\rm{DM}}=1.2\times10^{13} \ \rm{M_{\odot}}$ and $M_{\rm{DM}}= 1.0\times10^{13} \ \rm{M_{\odot}}$, respectively (Section \ref{sec:galaxies}). Defining the baryon mass fraction as $f_{\rm{b}}=M_{\rm{b,tot}}/(M_{\rm{DM}}+M_{\rm{b,tot}})$, we obtain $f_{\rm{b}}\approx0.11$ for NGC1961 and  $f_{\rm{b}}\approx0.09$ for NGC6753. 

The cosmic baryon mass fraction has been precisely measured by the \textit{Wilkinson Microwave Anisotropy Probe} (WMAP), yielding  $f_{\rm{b,WMAP}}=0.171\pm0.009$ \citep{dunkley09}. This value is significantly larger than that observed in NGC1961 and NGC6753, implying that $\sim$$30-50\%$ of the baryons are missing. This result is in agreement with observations of other disk galaxies, and galaxy groups \citep[see][for a review]{bregman07}. We note that a factor of $\sim2$ lower $M_{\rm{DM}}$ would imply a baryon mass fraction that is in agreement with the cosmic value. However, the quoted dark matter mass is broadly consistent with values found for massive spiral galaxies \citep{mandelbaum06}. Hence the missing baryon problem appears to be genuine. As a further caveat, we emphasize again the large uncertainties  associated with the determination of the hot gas mass within the virial radius. This plays a particularly important role given that $\sim$$50\%$ of the total baryon mass is in the form of hot gas in the (yet undetected) outer parts of the halo.   In principle, the real gas density profile beyond $\sim$$0.15r_{200}$ may significantly deviate from the extrapolated ones, yielding a very different total hot gas mass. For example, if the hot gas mass of NGC1961 and NGC6753 is  $2-3$ times more than that estimated based on the density profiles, the resulting $f_{\rm{b}}$ would be approximately consistent with the cosmic baryon mass fraction. To clearly determine the baryon budget of these (and other) spiral galaxies, further  deep X-ray observations are required, which probe their hot gas content  to larger radii.

\subsection{Iron bias and abundances}
\label{sec:ironbias}
The major parameters of the X-ray coronae, listed in Tables \ref{tab:values} and \ref{tab:luminosity}, are deduced from the best-fit spectra. However, two well-known complications are associated with the analysis of thermal X-ray spectra. One of them is the iron bias \citep[e.g.][]{buote00,bogdan12}, and the other is the degeneracy  between the emission measure and the abundance of the thermal component \citep[e.g.][]{david06}. 

In principle, the observed metal abundances may partly be influenced by the iron-bias, which appears if either a multi-temperature plasma or a temperature gradient is fit with a single temperature model. If the hot gas around NGC1961 and NGC6753 has a complex thermal structure, the measured metal abundances may be underestimated. The degeneracy between the emission measure and the abundances of the thermal model may have an impact on the observed  gas parameters. For example, if the abundance of the X-ray gas is not strictly sub-solar, but is about Solar, i.e. $\sim$$10$ times higher than measured, the gas densities, gas masses, and cooling times must be reduced by a factor of $\sim$$3.3$. 

To address these points, the thermal and abundance structure of the hot gas must be better explored. However, based on the presently available \textit{XMM-Newton} data set, the application of more complex (e.g. multi-temperature) models is not feasible. Therefore, additional deep X-ray observations of NGC1961 and NGC6753 are needed to settle this issue.

\subsection{Future prospects}
To further probe the luminous X-ray coronae and  constrain structure formation models, it is essential to identify X-ray coronae around other massive spiral galaxies. The large effective area of \textit{XMM-Newton} makes it especially suitable for the detection of hot gaseous coronae. Indeed, by combining the data of the three EPIC cameras, it is possible to collect a fairly large number of photons in moderately deep observations. Moreover, the large FOV allows one to explore the coronae out to sufficiently large radii. A disadvantage of \textit{XMM-Newton} is its relatively high instrumental background level, which results in notable systematic uncertainties, thereby hindering the study of faint X-ray emission. For example, in NGC1961 and NGC6753  the systematic  uncertainties dominate beyond $\sim$$60$ kpc radius, hence our understanding of their X-ray coronae would not  significantly improve at larger radii even with much deeper observations.   \textit{Chandra} has a lower level of instrumental background, making it ideal to explore the coronae at large galactocentric distances. However, due to the smaller effective area of  \textit{Chandra} it takes significantly longer to collect a meaningful number of photons. 

The upcoming \textit{eROSITA} mission will perform a four year all sky-survey program \citep{predehl10}, during which a mean exposure time of $2.0$ ks will be observed. This mission will play a major role in identifying the X-ray coronae of all nearby massive spiral galaxies. However, due to the relatively short exposure time of the \textit{eROSITA} survey, follow-up observations will be required to characterize the X-ray coronae, which can be done with \textit{Chandra}, \textit{XMM-Newton}, or \textit{eROSITA} itself. 

On a longer time scale, the \textit{Square Meter Arcsecond Resolution X-ray Telescope (SMART-X)} mission will offer an outstanding possibility to characterize hot X-ray coronae around massive spiral galaxies \citep{vikhlinin12}. With its large FOV ($22\arcmin\times22\arcmin$) and large effective area ($2.3 \ \rm{m^2}$ at $1$ keV) the \textit{SMART-X} Active Pixel Sensor Imager will collect $5500$ counts around NGC1961 in a $10$ ks observation in the $0.7-2$ keV energy range from the region between $25-45$ kpc. This will drastically increase the signal-to-noise ratio, thereby allowing the study of X-ray coronae in unprecedented details. Additionally, with the use of the X-ray Microcalorimeter Imaging Spectrometer aboard \textit{SMART-X} the outer regions of the X-ray coronae can also be explored. Indeed, with the microcalorimeter the soft Galactic emission can be differentiated from the hot X-ray gas corona. Thus, it will be feasible to investigate the X-ray coronae of a large sample of disk galaxies, thereby further constraining galaxy formation scenarios.

\section{Conclusions}
In this paper, we study the hot gaseous X-ray coronae around two ``normal'' spiral galaxies, NGC1961 and NGC6753, based on moderately deep \textit{XMM-Newton} observations. Since luminous X-ray coronae are a fundamental prediction of structure formation models, we use our observations to probe hydrodynamical structure formation simulations. Our results are:

\begin{enumerate}
\item We detect luminous X-ray coronae around NGC1961 \citep[in good agreement with][]{anderson11} and NGC6753. In both galaxies, the hot gas extends to $\sim$60 kpc, which significantly exceeds their optical radii. The spatial distribution of the hot gas is fairly round  and symmetric, indicating that the X-ray gas resides in hydrostatic equilibrium in the potential well of the galaxies. 
\item We find that the observed characteristics of the coronae in NGC6753 are fairly similar to NGC1961. In particular, within the  $(0.05-0.15)r_{200} $ region we measure gas temperatures of $kT\sim0.6$ keV and abundances of $\sim$$0.12$ Solar. The cooling time scale of the gas is $\sim$$50$ Gyrs, implying its quasi-static nature. The bolometric X-ray luminosity of the hot gas within the same region is $\sim$$6\times10^{40} \ \rm{erg \ s^{-1}}$ for both galaxies. 
\item We derive the baryon mass fraction of NGC1961 and NGC6753, and obtain  $f_{\rm{b}}\approx0.11$  and  $f_{\rm{b}}\approx0.09$, respectively. These values are lower than the cosmic baryon mass fraction, indicating that $\sim$$30-50\%$ of the baryons are missing from these galaxies. However,  large uncertainties are associated with the derivation of the total hot gas mass within the virial radius, which could bias the obtained values of $f_{\rm{b}}$.  
\item The observed characteristics of the X-ray coronae are compared with the moving mesh code \textsc{arepo} and the SPH-based code \textsc{gadget}. The implemented subresolution physics and the gravity solver are identical in the two codes, but they use different methods to solve the equations of hydrodynamics. We find that, while neither model gives a perfect description, the \textsc{arepo} code better reproduces  the luminosities, gas masses, and abundances observed around NGC1961 and NGC6753. Additionally, the shape of the observed density profiles are also in better agreement with the moving mesh code  within $\sim$$0.5r_{200}$. However, neither model incorporates efficient feedback from AGN and supernove, which could significantly alter the simulated properties of the coronae. 
\end{enumerate}

\begin{small}
\noindent
\textit{Acknowledgements.}
We thank the anonymous referee for useful comments. We thank Alexey Vikhlinin for helpful discussions about \textit{SMART-X}, Volker Springel for careful reading, and Rob Crain for critical comments. This work uses observations obtained with \textit{XMM-Newton}, an ESA science mission with instruments and contributions directly funded by ESA Member States and NASA. This publication makes use of data products from the Two Micron All Sky Survey, which is a joint project of the University of Massachusetts and the Infrared Processing and Analysis Center/California Institute of Technology, funded by the National Aeronautics and Space Administration and the National Science Foundation.  In this work the NASA/IPAC Extragalactic Database (NED) have been used. The authors acknowledge use of the HyperLeda database (http://leda.univ-lyon1.fr). \'AB acknowledges support provided by NASA through Einstein Postdoctoral Fellowship grant number PF1-120081 awarded by the Chandra X-ray Center, which is operated by the Smithsonian Astrophysical Observatory for NASA under contract NAS8-03060. WF and CJ acknowledge support from the Smithsonian Institution. 
\end{small}

\end{document}